\documentclass[12pt]{elsart}
\usepackage{feynmp,epsfig,graphics,color}
\def\sign{{\rm sign}\,}

\def\tr{{\rm tr} \,}

\def\pslash{\FMslash p}

\def\uslash{\FMslash u}

\def\wslash{\FMslash w}
\def\vslash{\FMslash v}

\def\w2{\tilde w^2}
\def\ws2{1}

\begin{document}
%\color{blue}
\begin{frontmatter}
\title{Antikaons and hyperons \\
in nuclear matter with saturation}
\author[GSI]{M.F.M. Lutz,}
\author[PTE]{C.L.\ Korpa}
\author[GSI]{and M.\ M\"oller}
\address[GSI]{Gesellschaft f\"ur Schwerionenforschung (GSI),\\
Planck Str. 1, 64291 Darmstadt, Germany}
\address[PTE]{Department of Theoretical Physics, University of
Pecs, \\Ifjusag u.\ 6, 7624 Pecs, Hungary}
%\today
\begin{abstract}
We evaluate the antikaon and hyperon spectral functions in a self-consistent and covariant many-body approach. The computation is
based on coupled-channel dynamics derived from the chiral SU(3)
Lagrangian. A novel subtraction scheme is developed
that avoids kinematical singularities and medium-induced power divergencies
all together. Scalar and vector mean fields are used to
model nuclear binding and saturation. The effect of the latter is striking for the
antikaon spectral function that becomes significantly more narrow at small momenta.
Attractive mass shifts of about 30 and 40 MeV are predicted for the
$\Lambda(1405)$ and $\Sigma(1385)$ resonances.
Once scalar and vector mean fields for the nucleon are switched on the $\Lambda(1520)$
resonances dissolves almost completely in nuclear matter.
All together only moderate attraction is predicted for the nuclear
antikaon systems at saturation density. However, at larger densities we predict a sizable population of
soft antikaon modes that arise from the coupling of the antikaon to a highly collective
$\Lambda(1115) $ nucleon-hole state. This may lead to the formation of exotic nuclear systems with strangeness and
antikaon condensation in compact stars at moderate densities.
\end{abstract}
\end{frontmatter}

%\tableofcontents

\section{Introduction}

The first attempts to predict the properties of antikaons in cold nuclear matter based on realistic
interactions are due to Waas, Kaiser and Weise \cite{Waas1,Waas2}. Starting from the
chiral SU(3) Lagrangian  the available low-energy antikaon nucleon
scattering data were fitted with s-wave amplitudes obtained from a phenomenological
coupled-channel approach \cite{Kaiser:Siegel:Weise:1995}. It was assumed that in-medium effects are
dominated by Pauli blocking effects which lead to a significant broadening and repulsive mass shift of
the $\Lambda (1405)$ resonance. The latter resonance couples strongly to the $\bar K N$ channel and is
therefore of utmost relevance for the nuclear antikaon dynamics.
The crucial importance of Pauli blocking for the properties of the $\Lambda (1405)$ resonance
in nuclear matter was  demonstrated  before by Koch using another schematic model
\cite{Koch:1994}. Later it was pointed out by one of the authors that the realistic treatment of the
many-body effects requires a self-consistent approach: a significantly reduced antikaon mass has
a strong influence on the $\Lambda(1405)$ mass \cite{Lutz:1998}. Based on the s-wave scattering amplitudes
of \cite{Kaiser:Siegel:Weise:1995} it was found that the $\Lambda(1405)$ mass is not pushed up to higher
masses \cite{Lutz:1998}.  This result was reproduced qualitatively by
Ramos and Oset \cite{Ramos:Oset:2000} applying a different coupled-channel model in a partially
self-consistent computation that was based on s-wave scattering exclusively. The computation \cite{Ramos:Oset:2000}
relied on an angle-average approximation and therefore the results are not fully self consistent.

The possible importance of p-wave interactions was pointed out in
\cite{Kolomeitsev:Voskresensky:Kaempfer:1995,Lutz:Kolomeitsev:Hirschegg:2000,Kolomeitsev:Voskresensky:2003}.
A computation based on a phenomenological meson-exchange
interaction was performed by Tolos, Ramos, Polls and Kuo
\cite{Tolos:Ramos:Polls:Kuo:2001} using a partially
self-consistent scheme that makes a quasi-particle ansatz for the
antikaon spectral function. These authors found a significant
influence of p-wave scattering. Unfortunately it was never
demonstrated whether the interaction used in
\cite{Tolos:Ramos:Polls:Kuo:2001} is compatible with available
differential scattering data. A further step towards a realistic
description of antikaon propagation in nuclear matter was taken by
two of the authors in \cite{Lutz:Korpa:2002}. Based on the chiral
coupled-channel theory developed in \cite{Lutz:Kolomeitsev:2002} a
fully self-consistent and covariant many-body approach was
established that considered s-, p- and d-wave scattering. The
underlying interaction was demonstrated to be compatible with
available low-energy pion-, kaon- and antikaon-nucleon
differential scattering data \cite{Lutz:Kolomeitsev:2002}. The
computation was based on the on-shell reduction scheme developed
in \cite{Lutz:Korpa:2002,Lutz:Kolomeitsev:2002}. The latter is
based on a covariant projector algebra that considers in
particular the proper mixing of partial waves in a nuclear
environment. It does not lead to any artifacts like a-causal
propagation if applied to the many-body system even in the
presence of p- and d-wave interactions. The results obtained in
\cite{Lutz:Korpa:2002} differ significantly from those of Tolos,
Ramos, Polls and Kuo \cite{Tolos:Ramos:Polls:Kuo:2001} and also
from the recent works by Oset and coworkers
\cite{Tolos:Oset:Ramos:2006,Kaskulov:Oset:2006}. In
\cite{Tolos:Oset:Ramos:2006}  an on-shell factorization for s-wave
scattering was assumed. It was pointed out that for p-wave
interactions the on-shell factorization turns invalid in nuclear
matter. An additional prescription was devised to treat the
in-medium p-wave phase space.

In their previous work \cite{Lutz:Korpa:2002} two of the authors demonstrated that the inclusion of
p-wave scattering leads to additional and significant attraction for the $\Lambda(1405)$
resonance. Moreover, attractive mass shifts for the p-wave and d-wave resonances $\Sigma(1385)$ and
$\Lambda(1520)$ were predicted. The d-wave resonance dissolves almost completely already at
saturation density.

The effect of using an in-medium modified pion propagator in the
pion-hyperon sub-systems was found to be of minor importance in the studies
\cite{Cieply,Korpa:Lutz:2005,Roth:Buballa:Wambach:2005}.
Contrasted results are claimed in
\cite{Ramos:Oset:2000,Tolos:Ramos:Polls:Kuo:2001,Tolos:Oset:Ramos:2006}. The differences may be
due in part to the use of different pion spectral distributions but also different
subthreshold transition amplitudes $\bar K N \to \pi \Sigma $, which
are a direct measure for the importance of pion-dressing effects. In this work we do not try to put further
light on those pending discrepancies.

It should be stressed that so far all realistic computations
\cite{Lutz:1998,Ramos:Oset:2000,Tolos:Ramos:Polls:Kuo:2001,Lutz:Korpa:2002,Lutz:Kolomeitsev:2002,Tolos:Oset:Ramos:2006,Cieply,Korpa:Lutz:2005,Roth:Buballa:Wambach:2005}
predict moderate attraction in the antikaon spectral function only. This does not appear to support the
strong-attraction scenario advocated by Akaishi and Yamazaki  \cite{Akaishi:Yamazaki:2002,Akaishi:Dote:Yamazaki:2005}.
Nevertheless, further improvements are desirable and possible. There are two main issues to elaborate on.
First, the recent re-measurement of the $K^-p$ scattering length by the DEAR collaboration
\cite{DEAR} is in conflict to some of the old bubble chamber $K^-p$ low-energy scattering data
\cite{Borasoy:Nissler:Weise:2005,Borasoy:Meissner:Nissler:2006}. One of the authors assures that this
challenge is present also in the more sophisticated approach of  \cite{Lutz:Kolomeitsev:2002}. At present
it appears impossible,
given the established coupled-channel approaches, to simultaneously describe the new accurate $K^-p$
scattering length \cite{DEAR} together with the low-energy scattering data. In this work we will focus
on the second issue: further improve the many-body approach based on  the
coupled-channel theory \cite{Lutz:Kolomeitsev:2002}.

It is the purpose of the present work to explore the effect of
nuclear binding and saturation on the antikaon and hyperon
properties in more detail within a self-consistent framework. This requires a significant extension
of the covariant many-body approach developed
in \cite{Lutz:Korpa:2002}. The first work addressing this issue is
due to Waas, Rho and Weise \cite{Waas2}, where it was claimed that
such effects are small and unimportant. Similar findings were
reported in
\cite{Ramos:Oset:2000,Tolos:Ramos:Polls:Kuo:2001,Tolos:Oset:Ramos:2006,Cieply}.
The latter results assume an attractive mean field potential of
about $50$ MeV for the nucleon relying on a non-relativistic
many-body approach. So far the possible importance of  large
scalar and vector nucleon mean fields has not been studied.
Furthermore we will investigate in this work the reliability of
the angle-average approximation applied in \cite{Ramos:Oset:2000,Tolos:Oset:Ramos:2006}.

The work is organized as follows. In section 2 and 3 the covariant many-body approach of
\cite{Lutz:Korpa:2002} is generalized for
the presence of scalar and vector mean fields of the nucleons. Section 4 introduces a novel
renormalization scheme for the in-medium meson-baryon loop functions that avoids the occurrence of
medium-induced power divergent terms as well as the occurrence of kinematical singularities.
This is a crucial issue once p-wave interactions are considered.
Numerical results are presented in section 5. The work closes with
section 6 giving a summary and conclusions.

The main findings of this work can be summarized as follows. The
use of an angle-average in the evaluation of the antikaon-nucleon loop
functions appears overall quite reliable, however, with some notable exceptions. The mass and
width shifts for the p-wave and d-wave hyperon states can not always be accurately computed relying
on the angle-average approximation. Scalar and vector mean fields have a strong impact on the
antikaon spectral function that becomes significantly more narrow at small momenta. It is
demonstrated that the latter can not be reproduced by assuming a weak scalar mean field for the nucleon. Only
the combined consideration of large scalar and vector mean fields has a significant impact on the nuclear
antikaon dynamics. The mean fields affect the hyperon resonances, with the exception of the $\Lambda(1520)$
resonance, only moderately. We consolidate our previous prediction that the $\Lambda(1405)$ and $\Sigma(1385)$
resonances experience sizeable attractive mass shifts in cold nuclear matter. The $\Lambda(1520)$ dissolves
almost completely already at saturation density.

All together only moderate attraction is predicted for the nuclear
antikaon systems at saturation density. At larger densities we predict a sizable population of
soft antikaon modes that arise from the coupling of the antikaon to the $\Lambda(1115) $ nucleon-hole state.
The latter is pushed down to smaller masses significantly by a level-level repulsion of the
$\Lambda(1115) $ nucleon-hole and antikaon mode. We speculate that this may lead to the formation of deeply
bound and exotic nuclear systems with strangeness and antikaon condensation in compact stars at moderate densities.

\section{Self-consistent dynamics for strangeness in nuclear matter}

In this section we generalize the self-consistent and relativistic many-body framework
established in \cite{Lutz:Korpa:2002}. We will keep this work self-contained recalling crucial elements
of the work \cite{Lutz:Korpa:2002}.

The free-space and on-shell antikaon-nucleon scattering amplitude is
\begin{eqnarray}
\langle \bar K^{j}(\bar q)\,N(\bar p)|\,T\,| \bar K^{i}(q)\,N(p) \rangle
&=&(2\pi)^4\,\delta^4(q+p-\bar q-\bar p )\,
\nonumber\\
&& \!\!\!\!\!\times \,\bar u(\bar p)\,
T^{ij}(\bar q,\bar p ; q,p)\,u(p) \,,
\label{on-shell-scattering}
\end{eqnarray}
where  $\delta^4(..)$ guarantees energy-momentum conservation and $u(p)$ is the
nucleon isospin-doublet spinor. Note also $\bar K =(K^-, \bar K^0)$.
The vacuum scattering amplitude is decomposed into its isospin channels
\begin{eqnarray}
&&T^{i j}(\bar q,\bar p \,; q,p)
= T^{(0)}(\bar k,k;w)\,P^{ij}_{(I=0)}+
T^{(1)}(\bar k,k;w)\,P^{ij}_{(I=1)}\;,
\nonumber\\
&& P^{ij}_{(I=0)}= \frac{1}{4}\,\Big( \delta^{ij}\,1
+ \big( \vec \tau\,\big)^{ij}\,\vec \tau \,\Big)\, , \quad
P^{ij}_{(I=1)}= \frac{1}{4}\,\Big( 3\,\delta^{ij}\,1-
\big(\vec \tau \,\big)^{ij}\,\vec \tau \,\Big)\;,
\label{}
\end{eqnarray}
where $q, p, \bar q, \bar p$ are the initial and final antikaon and nucleon 4-momenta and
\begin{eqnarray}
w = p+q = \bar p+\bar q\,,
\quad k= \half\,(p-q)\,,\quad
\bar k =\half\,(\bar p-\bar q)\,.
\label{def-moment}
\end{eqnarray}
In quantum field theory the scattering amplitudes $T^{(I)}$ follow as the
solution of the Bethe-Salpeter matrix equation
\begin{eqnarray}
T(\bar k ,k ;w ) &=& K(\bar k ,k ;w )
+\int \frac{d^4l}{(2\pi)^4}\,K(\bar k , l;w )\, G(l;w)\,T(l,k;w )\;,
\nonumber\\
G(l;w)&=&-i\,S({\textstyle
{1\over 2}}\,w+l)\,D({\textstyle {1\over 2}}\,w-l) \,,
\label{BS-eq}
\end{eqnarray}
in terms of the Bethe-Salpeter kernel $K(\bar k,k;w)$, the free space nucleon
propagator $S(p)=1/(\pslash-m_N+i\,\epsilon)$ and
kaon propagator $D(q)=1/(q^2-m_K^2+i\,\epsilon)$.

The antikaon-nucleon scattering process is readily generalized from the
vacuum to the nuclear matter case. In compact notation we write
\begin{eqnarray}
&& {\mathcal T} = {\mathcal K} + {\mathcal K} \cdot {\mathcal G} \cdot {\mathcal T}  \;,\quad
{\mathcal T} = {\mathcal T}(\bar k,k; w,u) \;, \quad {\mathcal G} = {\mathcal G} (l;w,u) \,,
\label{hatt}
\end{eqnarray}
where the  in-medium scattering amplitude ${\mathcal T}(\bar k,k;w,u)$ and the two-particle
propagator ${\mathcal G}(l;w,u)$ depend now on the 4-velocity $u_\mu$
characterizing the nuclear matter frame. For nuclear matter moving with a velocity
$\vec u$ one has
\begin{eqnarray}
u_\mu =\left(\frac{1}{\sqrt{1-\vec u\,^2/c^2}},\frac{\vec u/c}{\sqrt{1-\vec u\,^2/c^2}}\right)
\;, \quad u^2 =1\,.
\label{}
\end{eqnarray}
We emphasize that (\ref{hatt}) is properly defined from a Feynman diagrammatic point of view even in the case
where the in-medium scattering process is no longer well defined due to a broad antikaon spectral function.
In this work we do not consider medium modifications of the interaction kernel, i.e. we
approximate ${\mathcal K} = K$. We exclusively study the effect of an in-medium modified two-particle
propagator ${\mathcal G}$
\begin{eqnarray}
&& {\mathcal S}(p,u) = \frac{1}{\pslash-\Sigma_V\,\uslash -m_N+\Sigma_S + i\,\epsilon} + \Delta S(p,u)\,,
\nonumber\\
&& \Delta S (p,u) = 2\,\pi\,i\,\Theta \Big[p \cdot u-\Sigma_V  \Big]\,
\delta\Big[(p-\Sigma_V\,u)^2-(m_N-\Sigma_S)^2\Big]\,
\nonumber\\
&& \qquad \qquad \times \,\Big( \pslash- \Sigma_V\,\uslash +m_N -\Sigma_S\Big)\,\Theta \Big[k_F^2+p^2-(u\cdot p)^2\,\big]\,,
\nonumber\\
&& {\mathcal G}(l;w,u) = -i\,\frac{{\mathcal S}({\textstyle
{1\over 2}}\,w+l,u)}{({\textstyle {1\over 2}}\,w-l)^2-m_K^2-\Pi({\textstyle {1\over 2}}\,w-l,u)} \,,
\label{hatg}
\end{eqnarray}
where the Fermi momentum $k_F$ parameterizes the density $\rho$ of isospin-symmetric nuclear matter.
It holds
\begin{eqnarray}
\rho = -2\,\tr \,\gamma_0\,\int \frac{d^4p}{(2\pi)^4}\,i\,\Delta S(p,u)
= \frac{2\,k_F^3}{3\,\pi^2\,\sqrt{1-\vec u\,^2/c^2}}  \;.
\label{rho-u}
\end{eqnarray}
As an extension of our previous works we incorporate the effect of nuclear binding and saturation modelled in
terms of scalar and vector mean fields.
For the scalar and vector mean fields of the nucleon we use the simple
parametrization
\begin{eqnarray}
\Sigma_V = 290\, {\rm MeV}\,\frac{\rho}{\rho_0}\,, \qquad
\Sigma_S = 350\,{\rm MeV}\,\frac{\rho}{\rho_0} \,,
\label{def-mean-field}
\end{eqnarray}
a quite conservative estimate \cite{Serot:Walecka:1986,Brockmann:1990,Drukarev:1991,Finelli:2004,Fuchs:2004,Plohl:Fuchs:2006}.
It is emphasized
that scalar and vector mean fields
of the nucleon are not observable quantities. They are scheme dependent and serve as a phenomenological
tool to model nuclear binding and saturation effects in a manifest covariant manner.

\begin{figure}[t]
\begin{center}
\includegraphics[width=14cm,clip=true]{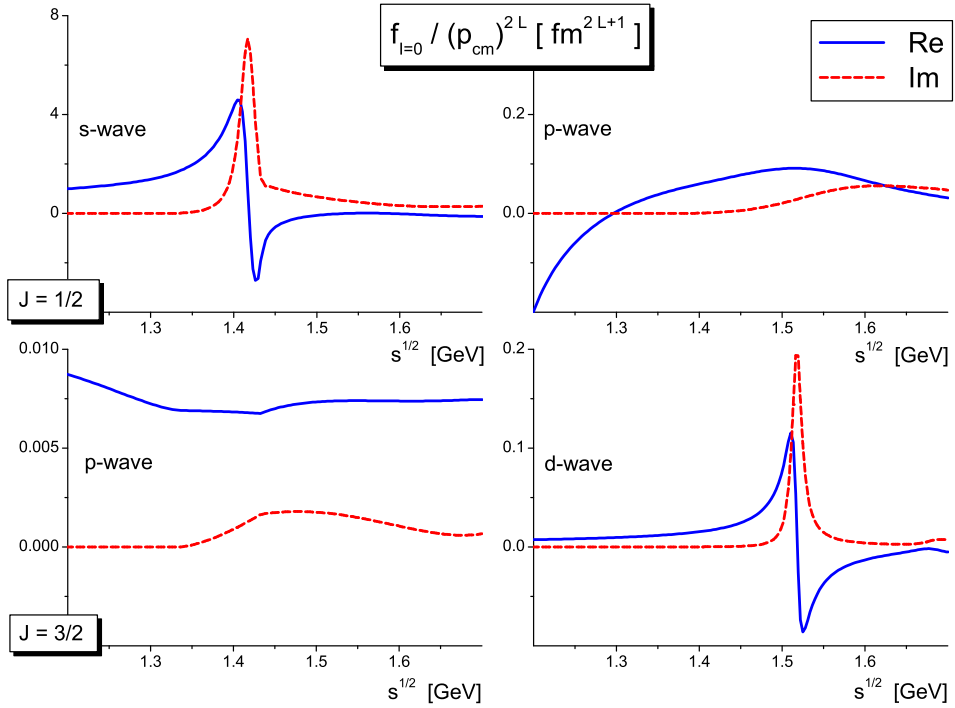}
\end{center}
\caption{Isospin-zero antikaon-nucleon scattering amplitudes with different angular momenta $L$ and $J$. The amplitudes
are taken from \cite{Lutz:Kolomeitsev:2002}. }
\label{fig:0a}
\end{figure}

\begin{figure}[t]
\begin{center}
\includegraphics[width=14cm,clip=true]{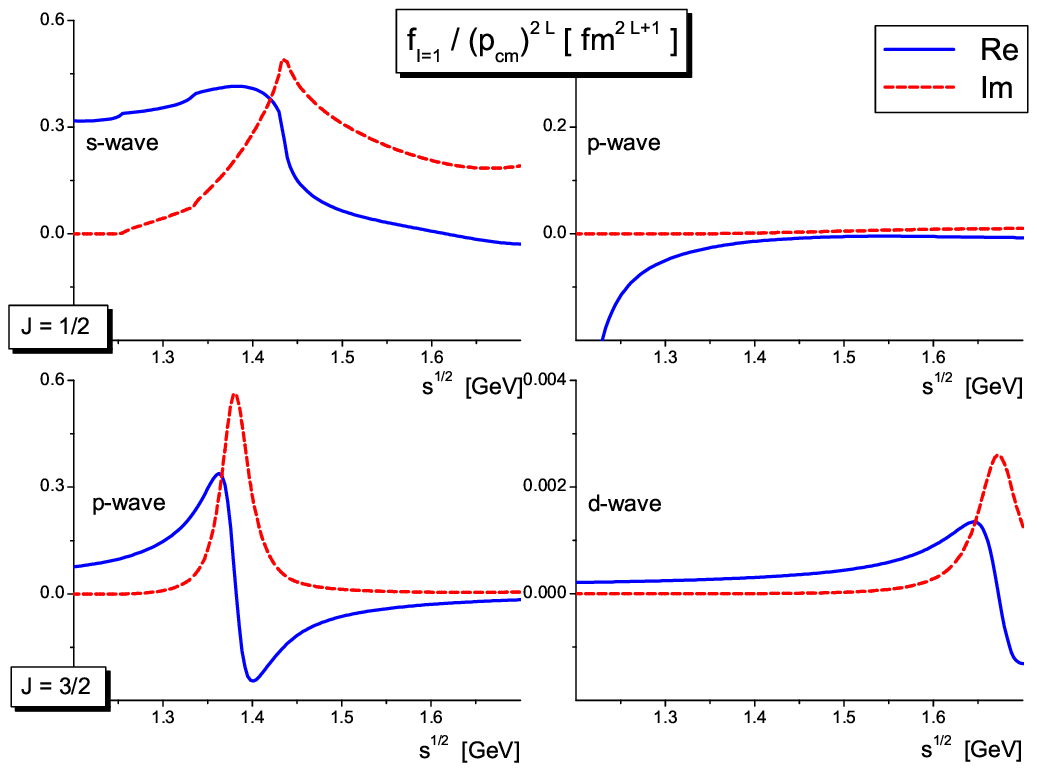}
\end{center}
\caption{Isospin-one antikaon-nucleon scattering amplitudes with different angular momenta $L$ and $J$. The amplitudes
are taken from \cite{Lutz:Kolomeitsev:2002}. }
\label{fig:0b}
\end{figure}

In the rest frame of the bulk matter with $u_\mu=(1,\vec 0\,)$ one recovers with (\ref{rho-u}) the
standard result $\rho = 2\,k_F^3/(3\,\pi^2)$. The antikaon self-energy $\Pi(q,u)$ is evaluated self-consistently in terms of the in-medium scattering amplitudes
${\mathcal T}^{(I)}(\bar k,k;w,u)$
\begin{eqnarray}
\Pi(q,u) &=& 2\,\tr \int \frac{d^4p}{(2\pi)^4}\,i\,\Delta S (p,u)\,
\bar {\mathcal T}\big({\textstyle{1\over 2}}\,(p-q),
{\textstyle{1\over 2}}\,(p-q);p+q,u \big)\,,
\nonumber\\
\bar {\mathcal T}&=& \frac{1}{4}\,{\mathcal T}^{(I=0)}+
\frac{3}{4}\,{\mathcal T}^{(I=1)} \;.
\label{k-self}
\end{eqnarray}
In order to solve the self-consistent set of equations (\ref{hatt},\ref{hatg},\ref{k-self})
it is convenient to rewrite the scattering amplitude as follows
\begin{eqnarray}
&&{\mathcal T}= K+K\cdot {\mathcal G}\cdot {\mathcal T}
= T+T\cdot \Delta {\mathcal G} \cdot {\mathcal T}\;,\quad
\Delta {\mathcal G}={\mathcal G}-G\;,
\label{rewrite}
\end{eqnarray}
where $T=K+K\cdot G\cdot T$  is the vacuum scattering amplitude. Given a set of tabulated
scattering amplitudes $T$ derived in free-space the self-consistent in-medium scattering amplitude
can be computed. The amplitudes used in this work are recalled from \cite{Lutz:Kolomeitsev:2002}.
For the readers convenience they are shown in Figs. \ref{fig:0a}-\ref{fig:0b}
in terms of conventional $f$ amplitudes as defined
\begin{eqnarray}
&& p_{\rm cm}\,f=  \frac{1}{2\,i}\,\Big( \eta \,e^{2\,i\,\delta}-1 \Big) \,, \qquad
\sqrt{s} = \sqrt{m_N^2+p_{\rm cm}^2}+\sqrt{m_K^2+p_{\rm cm}^2} \,.
 \label{def-f}
\end{eqnarray}
by the phase shift $\delta$ and inelasticity $\eta$.

\section{Covariant projector algebra }

Given the free-space scattering amplitude, $T$, of \cite{Lutz:Kolomeitsev:2002}
the solution of the self-consistent system (\ref{k-self}, \ref{rewrite})
is derived utilizing the projector algebra established in
\cite{Lutz:Korpa:2002}. The in-medium scattering amplitude, ${\mathcal T}$, takes the form
\begin{eqnarray}
&& {\mathcal T} =\sum_{i=1}^2 \sum_{j=1}^2 \,T^{(p)}_{[ij]}(v,u)\,P_{[ij]}(v,u)
\nonumber\\
&& \quad
+\sum_{i=1}^2 \sum_{j=3}^8 \,\Big(T^{(p)}_{[ij]}(v,u)\,P^\mu_{[ij]}(v,u)\,q_\mu
+T^{(p)}_{[ji]}(v,u)\,\bar q_\mu\,P^\mu_{[ji]}(v,u)\, \Big)
\nonumber\\
&& \quad + \sum_{i=3}^8 \sum_{j=3}^8\, T^{(p)}_{[ij]}(v,u)\,\bar
q_\mu\,P^{\mu \nu}_{[ij]}(v,u)\,q_\nu +\sum_{i=1}^2 \sum_{j=1}^2
T^{(q)}_{[ij]}(v,u)\,\bar q_\mu\,Q^{\mu \nu}_{[ij]}(v,u)\,q_\nu\,,
\nonumber\\
&& T^{(p)}(v,u)= M^{(p)}(v,u)\,\Big[ 1- \Delta J^{(p)}(v,u)\,M^{(p)}(v,u)
\Big]^{-1}\,,
\nonumber\\
&& T^{(q)}(v,u)= M^{(q)}(v,u)\,\Big[ 1- \Delta J^{(q)}(v,u)\,M^{(q)}(v,u)
\Big]^{-1}\,. \label{T-result}
\end{eqnarray}
The achievement of the representation (\ref{T-result}) lies in its similarity to a corresponding
expression obtained  previously in \cite{Lutz:Korpa:2002} for the limiting case of
vanishing vector mean field. In (\ref{T-result}) we introduced a convenient 4-momentum
\begin{eqnarray}
v_\mu = w_\mu - \Sigma_V\,u_\mu \,,
\end{eqnarray}
which we will be using throughout this work. The projectors
$P_{[ij]}(v,u)$ and $Q_{[ij]}(v,u)$ are recalled in Appendix A.
It holds
\begin{eqnarray}
&&P_{[ik]}\cdot P_{[lj]} =\delta_{kl}\,P_{[ij]} \;, \quad
P^\mu_{[ik]}\;\bar P^\nu_{[lj]}= \delta_{kl}\,P_{[ij]}^{\mu \nu}\,,\quad
\bar P^\mu_{[ik]}\,g_{\mu \nu}\,P^\nu_{[lj]}= \delta_{kl}\,P_{[ij]} \;,
\nonumber\\
&& Q_{[ik]}^{\mu \alpha }\,g_{\alpha \beta}\,P_{[lj]}^{\beta }
= 0 = \bar P_{[ik]}^{\alpha }\,g_{\alpha \beta}\,Q_{[lj]}^{\beta \nu }\;.
\label{proj-algebra-extension}
\end{eqnarray}
We mention that, due to the completeness of the projector algebra, an equivalent representation
of the in-medium scattering amplitude is possible in terms of the projectors $P_{[ij]}(w,u)$ and
$Q_{[ij]}(w,u)$ using the 4-momenta $w_\mu$ and $u_\mu$ rather than $v_\mu$ and $u_\mu$.

Before discussing in detail the matrix of loop functions $\Delta J_{[ij]}^{(p,q)}(v,u)$ we specify the
matrix of free-space scattering amplitudes $M^{(p,q)}_{[ij]}(v,u)$. Due
to the use of the projectors constructed in terms
of $v_\mu$ and $u_\mu$ this is slightly involved. The result (\ref{T-result}) is a consequence of the
representation
\begin{eqnarray}
&& T(\bar q, q;w)
=\sum_{i=1}^2 \sum_{j=1}^2 \,M^{(p)}_{[ij]}(v,u)\,P_{[ij]}(v,u)
\nonumber\\
&& \quad
+\sum_{i=1}^2 \sum_{j=3}^8 \,\Big(M^{(p)}_{[ij]}(v,u)\,P^\mu_{[ij]}(v,u)\,q_\mu
+M^{(p)}_{[ji]}(v,u)\,\bar q_\mu\,P^\mu_{[ji]}(v,u)\, \Big)
\nonumber\\
&& \quad + \sum_{i=3}^8 \sum_{j=3}^8\, M^{(p)}_{[ij]}(v,u)\,\bar
q_\mu\,P^{\mu \nu}_{[ij]}(v,u)\,q_\nu +\sum_{i=1}^2 \sum_{j=1}^2
M^{(q)}_{[ij]}(v,u)\,\bar q_\mu\,Q^{\mu \nu}_{[ij]}(v,u)\,q_\nu\,, \nonumber
\label{}
\end{eqnarray}
where the free-space amplitudes $M^{(p)}_{[ij]}(v,u)$ and $M^{(q)}_{[ij]}(v,u)$
are linear combinations of the $J^P=\frac{1}{2}^{\pm}, \frac{3}{2}^{\pm}$
partial wave amplitudes established in \cite{Lutz:Kolomeitsev:2002}. The latter are related to
the more conventional $f$ amplitudes of (\ref{def-f}) by
\begin{eqnarray}
 f_{J=L\pm 1/2} = \frac{p_{\rm cm}^{2\,J-1}}{8\,\pi \sqrt{s}}\,
\left( \frac{\sqrt{s}}{2}+ \frac{m_N^2-m_K^2}{2\,\sqrt{s}} \pm m_N\right)  M_{J^\pm} \,.
\label{relation-M-f}
\end{eqnarray}
More specifically we derive
\begin{eqnarray}
&&M^{(p)}_{[ij]}(v,u) = \sum_\pm \,C^{\frac{1}{2}^\pm}_{p,[ij]}(v,u)\, M_{\frac{1}{2}^\pm }(\sqrt{s}\,)
+\sum_\pm \,C^{\frac{3}{2}^\pm}_{p,[ij]}(v,u)\, M_{\frac{3}{2}^\pm }(\sqrt{s}\,)\,,
\nonumber\\
&&M^{(q)}_{[ij]}(v,u) =
\sum_\pm \,C^{\frac{3}{2}^\pm}_{q,[ij]}(v,u)\, M_{\frac{3}{2}^\pm }(\sqrt{s}\,)\,.
\label{recoupling-identity}
\end{eqnarray}
A complete list of the recoupling functions $C^{J^P}_{p,[ij]}(v,u)$ and $C^{J^P}_{q,[ij]}(v,u)$
is given in Appendix B.

The form of the loop functions can be taken over to a large extent from \cite{Lutz:Korpa:2002}.
Besides the generalization of \cite{Lutz:Korpa:2002} to the presence of scalar and vector mean fields
a few misprints are corrected. The reduced loop functions
$\Delta J_{[ij]}(v,u)$ acquire the generic form
\begin{eqnarray}
&&\Delta J_{[ij]}(v,u) = \int \frac{d^4l}{(2 \pi)^4}\,\Big[
g_{}(l;v,u)\,K^{}_{[ij]}(l;v,u)-g_{{\rm vac}}(l;v,u)\,K^{{\rm vac}}_{[ij]}(l;v,u)\Big]\,,
\nonumber\\ \nonumber\\
&& \;\;\;\;g_{}(l;v,u) =-\,\frac{i}{l^2-(m_N-\Sigma_S)^2+i\,\epsilon} \,
\frac{1}{(v-l)^2-m_K^2-\Pi(v-l,u)}
\nonumber\\
&& \qquad
+2\,\pi\,\Theta \Big(l\cdot u \Big)\,\delta(l^2-(m_N-\Sigma_S)^2)\,
\frac{\Theta \Big(k_F^2+(m_N-\Sigma_S)^2-(u \cdot l)^2 \Big)}{(v-l)^2-m_K^2-\Pi(v-l,u)}\,,
\nonumber\\
&& g_{\rm vac}(l;v,u) = \frac{-i}{(l+\Sigma_V\,u)^2-m_N^2+i\,\epsilon} \,\frac{1}{(v-l)^2-m_K^2+i\,\epsilon}\;,
\label{j-exp}
\end{eqnarray}
where the scalars $K_{[ij]}(l;v,u)$ and $K^{\rm vac}_{[ij]}(l;v,u)$ are linear in $m_N-\Sigma_S$ and
$m_N$ respectively. They involve powers of $l^2,l \cdot v$, $l \cdot u$ and $v \cdot u, v^2$. Detailed  results are
derived in the next section. The expressions (\ref{j-exp}) as they stand are ultraviolet
divergent. If regularized by a three momentum cutoff $\Lambda$, power divergent structures up to $\Lambda^4$ would
arise. This is unphysical and requires special attention. In the subsequent section
a renormalization scheme is introduced that eliminates all power divergent structures systematically.

The antikaon self-energy is determined by the in-medium scattering amplitudes $\bar T^{(p)}_{[ij]}(v,u)$,
properly isospin averaged. In an arbitrary frame it holds
\begin{eqnarray}
&&\Pi(q,u) =-\sum_{i,j=1}^8\,\int_0^{k_F} \frac{d^3 p}{(2\pi)^3}
\, \frac{2}{p_0}\,c^{(p)}_{[ij]}(q;w,u)\,\bar T^{(p)}_{[ij]}(w,u)
\nonumber\\
&& \qquad \qquad \;-\sum_{i,j=1}^2\,\int_0^{k_F} \frac{d^3
p}{(2\pi)^3} \, \frac{2}{p_0}\,c^{(q)}_{[ij]}(q;w,u)\,\bar
T^{(q)}_{[ij]}(w,u) \,,
\nonumber\\
&&\bar T_{[ij]}(w,u)
=\frac{1}{4}\,T^{(I=0)}_{[ij]}(w,u)+ \frac{3}{4}\,T^{(I=1)}_{[ij]}(w,u)\,, \label{kaon-final}
\end{eqnarray}
where $w_\mu=(p_\mu+q_\mu)$ and $p_0=\sqrt{(m_N-\Sigma_S)^2+\vec p\,^2}$. The coefficient functions
$c^{(p,q)}_{[ij]}(q;w,u)$ are recalled in Appendix C.
With  (\ref{T-result}, \ref{kaon-final})  and (\ref{j-exp}) a self-consistent set of
equations that defines the antikaon self-energy in terms of the free-space antikaon-nucleon scattering
amplitudes is derived. Given the partial-wave amplitudes $M_{J^P}(\sqrt{s}\,)$ together with a renormalization
scheme for the in-medium part of the loop function $\Delta J_{[ij]}(v,u)$
the antikaon self-energy can be computed numerically by iteration.

\newpage

\section{Computation of loop functions}

The evaluation of the real parts of the loop functions requires
great care. Consider the complete in-medium expressions
\begin{eqnarray}
&&J_{[ij]}(v,u) = \int \frac{d^4l}{(2 \pi)^4}\,
g_{}(l;v,u)\,K^{}_{[ij]}(l;v,u)\,,
\label{full-loop}
\end{eqnarray}
where we use the notation of (\ref{j-exp}).
It is assured that all diagonal loop functions
$J_{[ii]}(v,u)$ have positive imaginary parts everywhere as expected from causality. This
is an important consistency check of the projector approach defining the on-shell reduction
scheme \cite{Lutz:Kolomeitsev:2002,Lutz:Korpa:2002}.

A considerable simplification follows upon exploiting the explicit form of the projectors. They imply
that the matrix of loop functions can be composed out of 13 master loop functions $J_{i}(v,u)$ defined by
\begin{eqnarray}
&& J_i(v,u) =\int \frac{d^4l}{(2 \pi)^4}\,
g_{}(l;v,u)\,K^{}_{i}(l;v,u) \,,
\nonumber\\
&& K_0 = 1 \,, \qquad K_1 = \frac{l \cdot v}{\sqrt{v^2}}\,, \qquad K_2 = \frac{v^2\,(l \cdot u)-(v \cdot u)\,(v \cdot l)}{\sqrt{(v \cdot u)^2-v^2}\,\sqrt{v^2}}\,,
\nonumber\\
&& K_3 = {\textstyle { 1\over 2 }}\,\big[l^2- K_1^2+K_2^2  \big] \,, \qquad \! K_4 = K_1^2\,,\qquad \!
K_5 = K_2^2 \,,\qquad  \! K_6= K_1\,K_2\,,
\nonumber\\
&&   K_7 = K_1\,K_3\,, \qquad
K_8 = K_2\,K_3\,, \qquad K_9 = K_1^3 \,, \qquad  K_{10} = K_1^2\,K_2 \,, \qquad
\nonumber\\
&& K_{11} = K_2^3 \,, \qquad K_{12} =K_1\,K_2^2\,.
\label{def-Kis}
\end{eqnarray}
The matrix of loop functions $J_{[ij]}(v,u)$ of (\ref{full-loop}) is detailed
in Appendix D solely in terms of  linear combinations of the 13 master loop functions $J_i(v,u)$ as
introduced in (\ref{def-Kis}). We note that the latter decomposition defines implicitly the bare kernels
$K_{[ij]}(l;v,u)$ of (\ref{full-loop}). It is emphasized that it suffices to renormalize the 13
master loop functions.

The imaginary parts of the loop functions behave like $v_0^n$ for
large $v_0$ with $n $ not always smaller or equal to zero. Thus
power divergencies arise if the real parts are evaluated by means
of an unsubtracted dispersion-integral ansatz. The task is to
device a subtraction scheme that eliminates all such power
divergent terms. The latter are unphysical and in a
consistent effective field theory approach must be absorbed into
counter terms. Only the residual strength of the counter terms may
be estimated by a naturalness assumption reliably. Since we want
to neglect such counter terms it is crucial to set up the
renormalization in a proper manner, i.e. the residual counter terms should be
finite and of natural size. Only then it is justified to neglect the latter.

One may suggest to introduce a subtraction scheme in which the complete in-medium loop
functions approach in the zero-density limit the free-space form of the loop functions
as introduced in \cite{Lutz:Kolomeitsev:2002}. This would imply the representation
\begin{eqnarray}
&& J_{i}(v,u)\to_{\rho=0} N^{}_{i}(v)\,
\int_{-\infty}^{+\infty} \frac{d \,\bar v^2}{ \pi } \,
\frac{v^2}{\bar v^2}\frac{\rho(\bar v)}{\bar v^2-v^2-
i\,\epsilon } \,, \nonumber\\
&& \rho (v ) = \frac{\Theta \Big[v^2- (m_N+m_K)^2\Big]}{16\,\pi
\,\sqrt{v^2}}\,\sqrt{v^2
-2\,(m_N^2+m_K^2)+\frac{(m_N^2-m_K^2)^2}{v^2}}\,,
\label{impose}
\end{eqnarray}
where $N^{}_{i}(v)$ are kinematic functions of $\sqrt{v^2}$. The
latter are specified with
\begin{eqnarray}
&&N_{0}=1\;,\qquad \;\, \, N_{1}=\frac{v^2+m_N^2-m_K^2}{2\,\sqrt{v^2}}\,,
\qquad \,\,\,N_{2}=N_6=N_8=0\,,\qquad \,\,
\nonumber\\
&& N_{3}=-N_5=-\frac{[(m_N-m_K)^{2}-v^{2}]\,[(m_N+m_K)^{2}
-v^{2}]}{12\,v^{2}}\,,\qquad
N_{4}=N_1^2\,,
 \nonumber\\
&& N_{7}=-N_{12}=N_1\,N_3\,\,,\qquad N_{9}=N_1^3\,,
 \qquad N_{10}=N_{11}=0\,.
\label{def-Nij}
\end{eqnarray}

The loop functions are finite in the
zero-density limit as defined by (\ref{impose}). We recall that the representation
(\ref{impose}) was motivated by properties of the loop functions
manifest within dimensional regularization
\cite{Lutz:Kolomeitsev:2002}. Its form follows from the Passarino Veltman representation
\cite{Passarino:Veltman:1979} supplemented by a subtraction of reduced tadpole contributions.
The condition (\ref{impose}) defines a subtraction procedure that avoids the occurrence of
power divergent terms.

However, there is a subtle point we need to address. Since the projectors
exhibit kinematical singularities at $v^2 =0$ and $v^2=(v \cdot u)^2$ the loop functions are
correlated at this point necessarily. If such correlations are ignored the antikaon self-energy would
suffer from artificial structures that are at odds
with causality. In contrast to our previous work \cite{Lutz:Korpa:2002} where scalar and vector mean fields
were not considered the trouble some point $v^2 = (w-\Sigma_V\,u)^2=0$ is within the domain of validity of the
present approach. Thus it is crucial to devise a renormalization scheme which defines
the loop functions in a manner consistent with such constraints. Sufficient and necessary conditions that the kinematical
singularities cancel are readily derived:
\begin{eqnarray}
&& J^R_1 + \frac{(v \cdot u)}{\sqrt{(v \cdot u)^2}}\,J^R_2 = {\mathcal O} \left( \sqrt{v^2 } \right)\,, \qquad
 J^R_7 + \frac{(v \cdot u)}{\sqrt{(v \cdot u)^2}}\,J^R_8 = {\mathcal O} \left( \sqrt{v^2 } \right)\,,
\nonumber\\
&& J^R_4+J^R_5+2\,\frac{(v \cdot u)}{\sqrt{(v \cdot u)^2}}\,J^R_6= {\mathcal O} \left( \sqrt{v^2 } \right)\,,
\nonumber\\
&&J^R_{10}+J^R_{11}+2\,\frac{(v \cdot u)}{\sqrt{(v \cdot u)^2}}\,J^R_{12}= {\mathcal O} \left( \sqrt{v^2 } \right)\,,
\nonumber\\
&&\frac{(v \cdot u)}{\sqrt{(v \cdot u)^2}}\,J^R_{9}+3\,J^R_{10}+J^R_{11}+3\,\frac{(v \cdot u)}{\sqrt{(v \cdot u)^2}}\,J^R_{12}= {\mathcal O} \left( v^2  \right)\,.
\label{light-cone-condition}
\end{eqnarray}
and
\begin{eqnarray}
J^R_2 = J_3^R+J_5^R =J_6^R =J_8^R =J^R_7+J_{12}^R =0  \qquad {\rm at} \quad v^2 = (v \cdot u)^2 \,.
\label{decoupling}
\end{eqnarray}
We note that the condition (\ref{decoupling}) ensures that partial waves carrying
different total angular momentum decouple at the
point $v^2 = (v \cdot u)^2$. In general, at $v^2 \neq (v \cdot u)^2$
that is no longer true due to the non-conservation of total angular momentum in a nuclear
environment \cite{Lutz:Korpa:2002}.

Inspecting the free-space limit (\ref{impose}) with (\ref{def-Nij}) it is immediate that
additional subtractions are required as to ensure the cancellation of kinematical singularities.
The request (\ref{impose}) is incompatible with (\ref{light-cone-condition}, \ref{decoupling}).
We generalize the renormalization condition (\ref{impose}) appropriately:
\begin{eqnarray}
&& J^{R}_{i}(v,u)\to_{\rho=0} J^V_i(v)\equiv N^{}_{i}(v)\,
\int_{-\infty}^{+\infty} \frac{d \,\bar v^2}{ \pi } \,
\frac{v^2}{\bar v^2}\frac{\rho(\bar v)}{\bar v^2-v^2-
i\,\epsilon }
\nonumber\\
&&  \quad  + \,\Delta^{(4)}_{i}(v)\,
\int_{-\infty}^{+\infty} \frac{d \,\bar v^2}{ \pi } \,
\left( \frac{v^2}{\bar v^2}\right)^2\rho(\bar v) + \,\Delta^{(6)}_{i}(v)\,
\int_{-\infty}^{+\infty} \frac{d \,\bar v^2}{ \pi } \,
\left( \frac{v^2}{\bar v^2}\right)^3\rho(\bar v)\,,
\label{impose-generalization}
\end{eqnarray}
where the additional subtractions  are invoked if and only if they are unavoidable.
The terms $\Delta^{(4)}_{i}(v)$ and $\Delta^{(6)}_{i}(v)$ are detailed with
\begin{eqnarray}
&&\Delta^{(4)}_{3} =-\Delta^{(4)}_{5} = \frac{(m_N^2-m_K^2)^2}{3\,(v^2)^2} \,, \qquad
\Delta^{(4)}_{9} = -\frac{ (m_N^2-m_K^2)^3}{8\,\sqrt{v^2}\,(v^2)^2}\,,
\nonumber\\
&&\Delta^{(4)}_{7} =-\Delta^{(4)}_{12} = -\left(\frac{N_{1}}{12}
-\frac{N_{1}\,N_{5}}{v^2}-\frac{(m_{N}^2-m_K^2)^2}{8\,\sqrt{v^2}\,v^2}\right) \,,
\nonumber\\
&& \Delta^{(6)}_{7} =-\Delta^{(6)}_{12} = \frac{(m_N^2-m_K^2)^2\,N_{1}}
{12\,(v^2)^2}+\frac{(m_{N}^2-m_{K}^2)^2}{8\,\sqrt{v^2}\,v^2}+\frac{(m_{N}^2-m_{K}^2)^3}{24\,\sqrt{v^2}\,(v^2)^2}\,,
\end{eqnarray}
where we provide only those which are non-zero.

For the renormalized  loop functions we impose a dispersion-integral representation
in terms of spectral weight functions, $\Im J^{R}_{i}(\bar v_0;v_0,\vec w\,)$, that depend on 'external' and
'internal' energies $v_0=w_0-\Sigma_V$ and $\bar v_0$. For clarity of the presentation
we proceed in the rest frame of nuclear matter with $\vec u=0$.
We introduce the renormalized loop functions,
$J^R_{i}(v_0,\vec w\,)$,  as follows
\begin{eqnarray}
&&J^R_{i}(v_0,\vec w\,) = \int_{-\infty}^{+\infty} \frac{d \bar
v_0}{\pi} \left(\frac{\Im J^R_{i}(\bar v_0;v_0,\vec
w\,)}{\bar v_0-v_0-i\,\epsilon\,(\bar
v_0 -\mu)}  \right)\sign(\bar
v_0-\mu) + J^C_{i}(v_0, \vec w\,)
\nonumber\\
&& \qquad \quad \mu = \sqrt{(m_N-\Sigma_S)^2+k_F^2} \,,
\label{disp-integral}
\end{eqnarray}
with
\begin{eqnarray}
&&\Im J^R_{i}(\bar v_0; v_0,\vec w\,) = \int \frac{d\,^3
l}{2\,(2\, \pi)^3}\, \,\Big((m_N-\Sigma_S)^2+\vec l\,^2\,
\Big)^{-\frac{1}{2}}\,
\nonumber\\
&&  \qquad \times \Big\{ K^R_{i}(l_+,\bar v_0; v_0,\vec w\,)\,
\rho_K(\bar{v}_+, \vec w-\vec l\;)\, \Big[  \Theta(+\bar{v}_+
)-\Theta(k_F- | \vec l\,| ) \Big]
\nonumber\\
&&  \qquad\;+ \,K^R_{i}(l_-,\bar v_{0}; v_0,\vec
w\,)\,\rho_K(\bar{v}_-, \vec w-\vec l\;)\; \Theta(-\bar{v}_-
)\, \Big\} \,,
\nonumber\\
&& l_\pm^\mu = (\pm \,\sqrt{(m_N-\Sigma_S)^2+\vec l\,^2}, \vec l \;)\,,
\qquad \bar{v}_\pm = \bar{v}_0  \mp \sqrt{(m_N-\Sigma_S)^2+\vec l\,^2} \,,
\label{def-loop-J_n}
\end{eqnarray}
the antikaon spectral function
\begin{eqnarray}
\rho_K(\omega, \vec q\,) = - \frac{1}{\pi}\,\Im\, \frac{1}{\omega^2-\vec q\,^2-m_K^2-\Pi(\omega, \vec q\,)+i\,\epsilon}\,,
\label{def-kaon-spectral-function}
\end{eqnarray}
and scalar functions $K^{R}_{i}(l_+,\bar v_0; v_0,\vec w\,)$ that are of kinematic origin. The latter are listed in
Appendix E. In the limit $\bar v_0=v_0$ and $l^\mu_\pm = l^\mu$ they reproduce the
corresponding functions $K_{i}(l;v,u)$ introduced in (\ref{def-Kis}). It remains to specify the subtraction
terms, $J^C_i(v_0,\vec w\,)$, in (\ref{disp-integral}). In Appendix F they are defined in terms of the
integrals
\begin{eqnarray}
&&\bar C_{a,n}^{ijk}(\vec w\,) = \int_{-\infty}^{+\infty} \frac{d \bar
v_0}{\pi} \int \frac{d\,^3
l}{2\,(2\, \pi)^3}\, \,\Big((m_N-\Sigma_S)^2+\vec l\,^2\,
\Big)^{-\frac{1}{2}}\,
\nonumber\\
&&  \quad \!\! \!\times \Big\{
(\bar v \cdot u)^a\,\frac{(\bar l_+ \cdot \bar v)^i\,(\bar l_+ \cdot u)^j\,(\bar l^{\,2}_+\,)^k}{(\bar v^2)^n}\,
\rho_{K}(\bar{v}_+, \vec w-\vec l\;)\, \Big[  \Theta(+\bar{v}_+
)-\Theta(k_F- | \vec l\,| ) \Big]
\nonumber\\
&&  \quad \!+ \,(\bar v \cdot u)^a\,\frac{(\bar l_- \cdot \bar v)^i\,(\bar l_- \cdot u)^j\,(\bar l^{\,2}_-\,)^k}{(\bar v^2)^n}\,
\rho_{K}(\bar{v}_-, \vec w- \vec l\;)\; \Theta(-\bar{v}_-
)\, \Big\} \,,
\nonumber\\
&& \bar l^\mu_\pm = l^\mu_\pm - \frac{1}{2}\,\bar v^\mu  \,, \qquad  \bar v^2 = (\bar
v \cdot u)^2 - (v \cdot u)^2 +v^2 \,, \qquad u^\mu=(1, \vec 0\,)\,,
\label{def-loop-C_n}
\end{eqnarray}
where the notations of (\ref{def-loop-J_n}) are applied.
We assure that the representation (\ref{disp-integral}) is compatible with the constraints
(\ref{light-cone-condition}, \ref{decoupling}, \ref{impose-generalization}), i.e. kinematical singularities
are avoided. It is noted that (\ref{light-cone-condition}) would hold even for $J^C_i \to 0$. Non-vanishing
subtraction terms $J^C_i\neq 0$ are required as to guarantee the decoupling of partial waves at vanishing
three momentum $\vec w$ as well as consistency with the free-space limit (\ref{impose-generalization}).

The important achievement of the representation (\ref{def-loop-J_n}) lies in the asymptotic properties of
$\bar v_0\,\Im J^R_{i}(\bar v_0,v_0,\vec w\,)$  for large $\bar v_0$.
By construction they are bounded functions, which guarantees that the dispersion integrals in (\ref{disp-integral})
are finite. Here we make the physical assumptions that at large energies or large momenta the in-medium self-energy of the antikaon approaches zero. This is equivalent to assuming that the
antikaon spectral function is normalized to its canonical free-space value.
We point out that in addition all coefficients $\bar  C_{a,n}^{ijk}$ that occur in the evaluation of $J_i^C$ are finite
as well. This follows from the expressions in Appendix F and the
asymptotic behavior of the integrand of (\ref{def-loop-C_n}) for large $\bar v_0$. For instance at
$\vec w=\vec u= 0$ it holds in this limit
\begin{eqnarray}
&& \bar l_\pm \cdot \bar v \to \frac{m_N^2-m_K^2}{2} \,, \qquad \qquad \quad
\bar l_\pm^{\,2} \to \frac{m_N^2+m_K^2}{2} -\frac{\bar v_0^2}{4} \,, \quad
\nonumber\\
&& \bar l_\pm \cdot u \to \frac{m_N^2-m_K^2}{2\,\bar v_0}\,.
\label{asymptotic}
\end{eqnarray}
As a consequence of (\ref{asymptotic}) the integrals of (\ref{def-loop-C_n}) are finite for
\begin{eqnarray}
&& a-j+2\,k-2\,n \leq -2 \,\qquad {\rm if} \qquad a+j ={\rm odd} \,,
\nonumber\\
&& a-j+2\,k-2\,n \leq -1 \,\qquad {\rm if} \qquad a+j ={\rm even} \,.
\label{condition-finite}
\end{eqnarray}
Upon inspection of Appendix F, indeed, there occur only such coefficients, $\bar C_{a,n}^{ijk}$ with $a,n,j,k$ compatible
with (\ref{condition-finite}).

There is yet another important issue to be discussed that is related to the evaluation of the loop functions
(\ref{disp-integral}). For $n> 1$ the integral in (\ref{def-loop-C_n}) is not always defined properly. The integral over
$\bar v_0$ may be ill behaved at $\bar v_0 = \pm |\vec w \,|$ in this case. It is emphasized that this is in contrast to
the integrals of (\ref{disp-integral}). Due to the particular structure of the kernels $K_i^R$ the latter are
finite always upon the application of  the principal value prescription. It is noted that in the free-space limit
all coefficients $\bar C_{a,n}^{ijk}(\vec w\,)$ approach a constant. This
property is a direct consequence of covariance. Moreover, at finite density the $\bar v_0$-integral of (\ref{def-loop-C_n})
is well defined for sufficiently small three momenta $\vec w$.
In this case the troublesome region $\bar v_0 \sim \pm |\vec w|$ is excluded as can be verified by a phase space argument.
Since the antikaon spectral function of (\ref{def-kaon-spectral-function})
is non-zero for
\begin{eqnarray}
\omega >\omega^+_{thr}= m_\Lambda -\sqrt{(m_N-\Sigma_S)^2+k_F^2}-\Sigma_V\,, \qquad
\omega < \omega^-_{thr}=- m_K \,,
\end{eqnarray}
only, the critical condition reads
\begin{eqnarray}
| \vec w \,| < |\omega^+_{thr}|+\sqrt{(m_N-\Sigma_S)^2+k_F^2} \,,
\label{w-crit}
\end{eqnarray}
where we used $|\omega^-| > |\omega^+|$. Thus we may introduce well behaved coefficients by a Taylor expansion,
\begin{eqnarray}
C_{a,n}^{ijk}(\vec w\,) = \bar C_{a,n}^{ijk}(0) +
\frac{1}{2}\,\vec w^2\,(\nabla_{\vec w} \cdot \nabla_{\vec w})\,\bar C_{a,n}^{ijk}(0) \,,
\label{renormalized-Cs}
\end{eqnarray}
where we keep the minimal order as to ensure consistency with (\ref{decoupling}, \ref{impose-generalization}).
In Appendix G the counter loops $J_i^C(v_0,\vec w)$ of (\ref{disp-integral}) are expressed in terms of
the coefficients  (\ref{renormalized-Cs}).
With (\ref{renormalized-Cs}) the specification of the functions $J_i^R(v_0,\vec w)$ is completed.
The renormalized form, $J^R_{[ij]}(v_0,\vec w\,)$, of the  full loop matrix
in (\ref{full-loop}) is given by $J^R_{i}(v_0,\vec w\,)$ in terms of the linear algebra of Appendix D.

The renormalized form of the in-medium part of the loop functions (\ref{j-exp}) is decomposed with
\begin{eqnarray}
\Delta J_{[ij]}(v_0,\vec w\,) = J^R_{[ij]}(v_0,\vec w\,)
- J^V_{[ij]}(v_0,\vec w\,) -\Sigma_V\,\Delta J^V_{[ij]}(v_0,\vec w\,) \,,
\label{final-loop-renormalized}
\end{eqnarray}
where it is left to specify $J^V_{[ij]}(v_0,\vec w\,) $ and $\Delta J^V_{[ij]}(v_0,\vec w\,) $.
In Appendix G the latter are composed out of the 6 master functions
$J^V_{i}(w)$ with $i=0,1,3,4,7,9$ introduced already in (\ref{impose-generalization}).

It should be emphasized that the antikaon self-energy $\Pi(\omega, \vec q\,)$ as given by (\ref{kaon-final}) should
be trusted only for positive energy, i.e. where the self-energy describes the propagation properties of antikaons in
our convention. For negative energies, $\omega <0$, where the self-energy determines the properties of kaons
we approximate the self-energy in (\ref{disp-integral}) by a energy and momentum independent term linear in the
density. The latter constant is adjusted as to reproduce the well established repulsive kaon mass
shift of about 20 MeV at saturation density. Our numerical simulations reveal that
the effect of including or excluding this effect is of very minor importance.

We close this section with a brief exposition of the angle-average approximation applied in
\cite{Ramos:Oset:2000,Tolos:Oset:Ramos:2006}. The loop functions as introduced in our work are
frame independent, being function of the two scalars $v^2$ and $v \cdot u$ only. Thus we may evaluate the
loop functions in any frame. In order to connect to the angle-average approximation it is necessary to
compute the imaginary part of the loop functions in the center of mass frame with
\begin{eqnarray}
\bar v^{cm}_\mu=(\frac{\bar v^{}_0}{| \bar v^{}_0|}\,\sqrt{\bar v^2},\vec 0\,)\,,
\label{cm-frame}
\end{eqnarray}
where we use an upper script 'cm'
to make clear in which frame we are. In the rest frame of nuclear matter we have
\begin{eqnarray}
\bar v_\mu=(\bar v_0, \vec w\,)\,, \qquad  \qquad u_\mu=(1,\vec 0)\,.
\label{rest-frame}
\end{eqnarray}
The 4-velocity of nuclear matter as given in the center of mass frame is readily identified
\begin{eqnarray}
\bar v \cdot u= \bar v^{cm} \cdot u^{cm} \,, \qquad
u^{cm}_\mu= (\sqrt{1+\vec u\,^2},\vec u\,) \,,\qquad |\vec u\,|= \frac{|\vec w\,|}{\sqrt{\bar v^2}}\,.
\label{identify}
\end{eqnarray}
Boosting into the center of mass frame  we
derive the representation
\begin{eqnarray}
&&\Im J^R_{i}(v_0,\bar{v}_0,\vec w\,) = \int \frac{d\,^3
l}{2\,(2\, \pi)^3}\, \,\Big((m_N-\Sigma_S)^2+\vec l\,^2\,
\Big)^{-\frac{1}{2}}\,
\nonumber\\
&&  \qquad \times \Big\{ K^R_{i}(l_+,v,\bar v,u\,)\,
\rho_K(\bar{v}_+, \vec k_+\;)\, \Big[  \Theta(+\bar{v}_+
)-\Theta(k_F- | \vec k\,| ) \Big]
\nonumber\\
&&  \qquad\;+ \,K^R_{i}(l_-,v,\bar v,
u\,)\,\rho_K(\bar{v}_-, \vec k_-\;)\; \Theta(-\bar{v}_-
)\, \Big\} \,,
\nonumber\\
&& l_\pm^\mu = (\pm \,\sqrt{(m_N-\Sigma_S)^2+\vec l\,^2}, \vec l \;)\,,
\qquad
\bar{v}_\pm = \sqrt{1+\vec u\,^2}\,\big(\bar v_0^{cm}  -l_0^\pm \big)\,
+ \vec l \cdot \vec u \,,
\nonumber\\
&&\vec k\,^2_\pm = \vec l\,^2+ (\vec l \cdot \vec u\,)^2 + \vec u\,^2 \,(\bar v_0^{cm}-l_0^\pm)^2
+2\,(\vec l \cdot \vec u\,)\,(\bar v_0^{cm}-l_0^\pm)\,\sqrt{1+ \vec u\,^2}\,,
\nonumber\\
&& \vec k\,^2\,= \vec l\,^2 + (\vec l\cdot \vec u\,)^2+\vec u\,^2\,(l_0^+)^2
-2\,(\vec l \cdot \vec u\,)\, l_0^+\,\sqrt{1+ \vec u\,^2}\,.
\label{def-loop-J_n-alt}
\end{eqnarray}
By explicit numerical simulations we confirm that (\ref{def-loop-J_n}) and (\ref{def-loop-J_n-alt}) agree
identically. Consequently also the real part of the loop functions defined by (\ref{disp-integral})
coincide. In the numerical result section we will provide detailed comparisons of full simulations with those
relying on an angle-average approximation:  in the center of mass frame we take the angle-average of
$\vec k_\pm^2 $ and $\bar v_\pm $ in (\ref{def-loop-J_n-alt}). In addition we  assume the factorization
\begin{eqnarray}
&&\int \frac{d \Omega_l}{4 \pi }\, K^R_i ( \vec l , \vec u\,) \,\Theta (k_F^2-\vec k^2)
\nonumber\\
 \qquad \to && \qquad
\left(\int \frac{d \Omega_l}{4 \pi }\, K^R_i ( \vec l,  \vec u\,)\right) \left(
\int \frac{d \Omega_l}{4 \pi }\,\Theta (k_F^2-\vec k^2) \right)\,.
\label{factorization}
\end{eqnarray}
A corresponding approximation is applied to (\ref{def-loop-C_n}). The free-space loop matrix $J^V_{[ij]}$ is
unchanged. These assumptions simplify the numerical simulations dramatically. The angle-average can be performed
analytically and it remains a one-dimensional integral only that needs to be evaluated numerically
\footnote{The angle-average approximation of \cite{Ramos:Oset:2000,Tolos:Oset:Ramos:2006}
is introduced with $\vec l \cdot \vec u \to 0$ in the expressions for
$\vec k\,^2_\pm$ and $\bar v_\pm$ but keeping the proper angle dependence
in the Pauli-blocking term $\Theta(k_F-| \vec k \,|)$ \cite{Tolos:Oset:Ramos:2006,private-Ramos-Tolos-2006}.
In addition the correction factor $\sqrt{1+\vec u\,^2}$ in $\bar v_\pm$ is
omitted \cite{Tolos:Oset:Ramos:2006,private-Ramos-Tolos-2006}.}.

It should be mentioned that such angle-average approximations have a long history in the nuclear
many-body literature. It causes a considerable simplification since it avoids the coupling of different partial
waves in nuclear matter. In the case of  the Bruckner-Hatree-Fock approach for the nuclear equation of state
it was proven to be a quite reliable approximation \cite{Jorth-Jensen:Kuo:Osnes:1995,Frick:Muther:2003}.
However, since the antikaon self-energy has a much more pronounced energy and momentum dependence it needs to
be checked whether this is also true for nuclear antikaon systems.

\newpage

\section{Numerical results}

\begin{figure}[b]
\begin{center}
\includegraphics[width=14cm,clip=true]{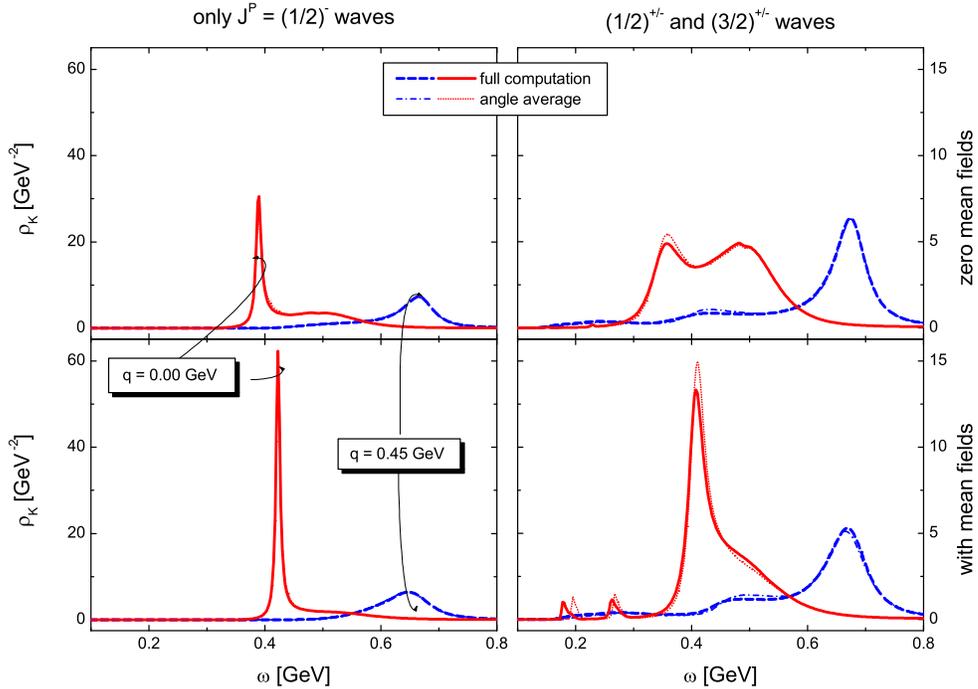}
\end{center}
\caption{Antikaon spectral function as a function of energy $\omega$ and momentum $\vec q$ at nuclear saturation
density. The first upper (lower) panel gives the results with switched off (on) mean fields. The left hand
panels consider  s-wave interactions only, whereas the right hand panels include the effects of
s-, p-, and d-waves. }
\label{fig:1}
\end{figure}

We briefly describe the numerical implementation the results are based on.
Throughout this section we assume
nuclear matter at rest, i.e. we put $u_\mu=(1, \vec 0\,)$. According to
the renormalization scheme described in great detail in the previous sections
there is, at least in principle, no need of any cutoff in the numerical simulation. However, since the
free-space scattering amplitudes are not available at all energies, the antikaon self-energy
can be evaluated only in a finite energy and momentum interval. The self-consistent system is
solved by iteration. At zeroth order the self-energy is computed in the $T \rho$ approximation
according to (\ref{kaon-final}) with the free-space scattering amplitude as given in
(\ref{recoupling-identity}) and \cite{Lutz:Kolomeitsev:2002}. The self-energy is computed for
$0< \omega < 1.4$ GeV and $0 < | \vec q \,|< 1.3$ GeV. In the next step the in-medium modification of
the loop functions $\Delta J^R_{[ij]}(v_0,\vec w\,)$ is evaluated using the renormalization scheme defined by
(\ref{disp-integral}, \ref{def-loop-J_n}). The quality of the angle-average approximation will be discussed below.
In the numerical simulation the in-medium
antikaon spectral function is put to its free-space limit outside the region where
the antikaon self-energy was computed. The iteration continues
by computing again the antikaon self-energy, however, now with the in-medium scattering amplitudes as
implied by (\ref{T-result}). The iteration
continues until convergence is reached. Typically this requires 4 to 5 iterations.

\begin{figure}[b]
\begin{center}
\includegraphics[width=12cm,clip=true]{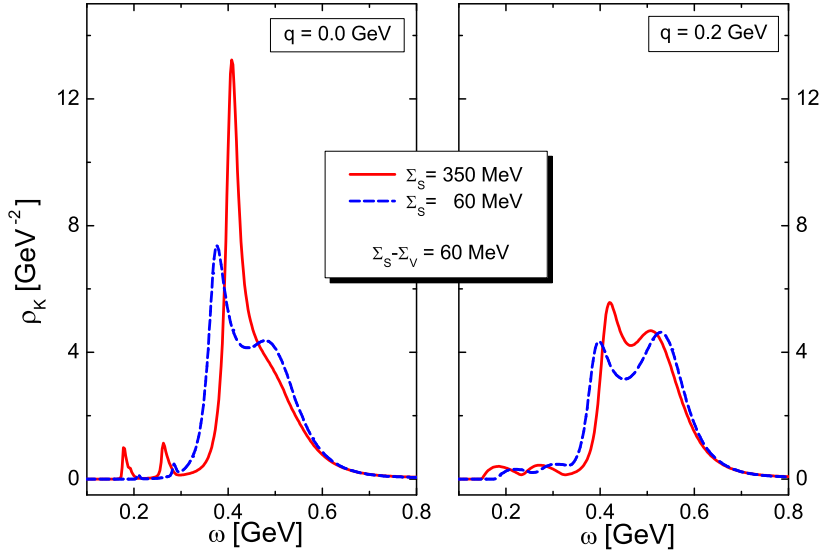}
\end{center}
\caption{Antikaon spectral function as a function of energy $\omega$ and momenta $\vec q=0$ and 200 MeV
at nuclear saturation density. The effects of s-, p-, and d-waves are considered. }
\label{fig:1b}
\end{figure}

\begin{figure}[t]
\begin{center}
\includegraphics[width=14.5cm,clip=true]{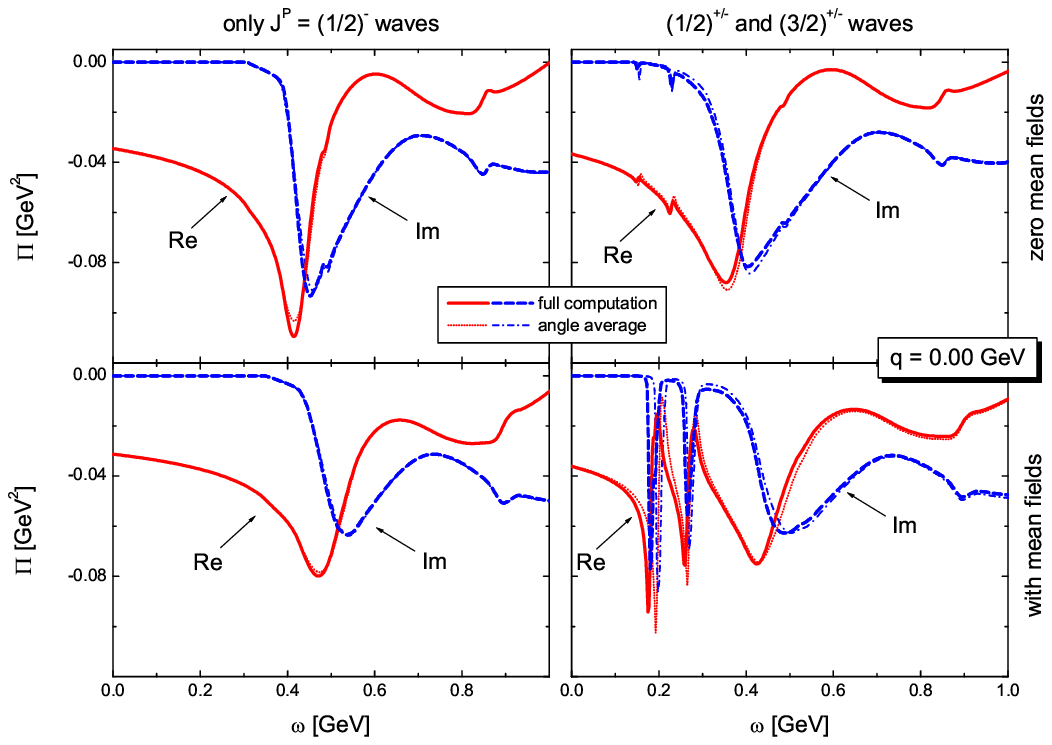}
\end{center}
\caption{Antikaon self-energy as a function of energy $\omega$ and momentum $\vec q$ at nuclear saturation
density. The first upper (lower) panel gives the results with switched off (on) mean fields. The left hand
panels consider  s-wave interactions only, whereas the right hand panels include the effects of
s-, p-, and d-waves. }
\label{fig:2}
\end{figure}

\begin{figure}[t]
\begin{center}
\includegraphics[width=14cm,clip=true]{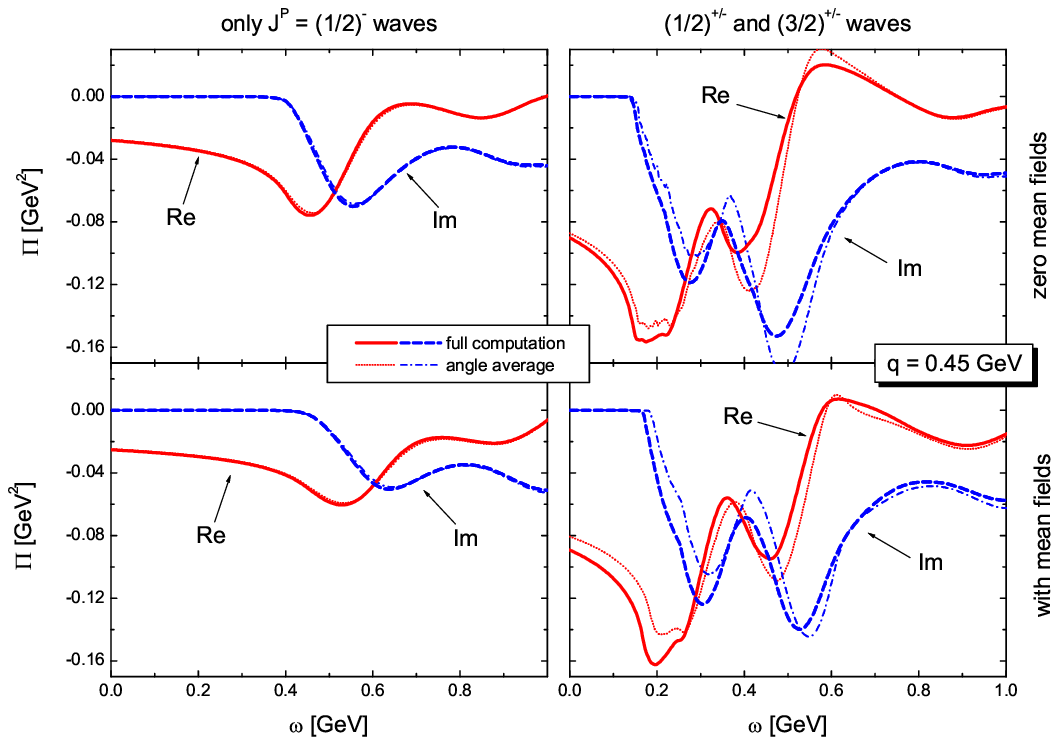}
\end{center}
\caption{ Same as Fig. 3 but for antikaon momentum of 0.45 GeV.}
\label{fig:3}
\end{figure}

In Fig. 1 we present our numerical results at nuclear saturation density for the antikaon spectral
function $\rho_K(\omega, \vec q\,)$ as a function of energy and momentum as defined in the rest frame of
nuclear matter. Results are shown for switched-on and switched-off scalar and vector mean fields.
As a reference we provide spectral functions that are based on s-wave interactions only.
Without scalar and vector mean fields our previous results \cite{Lutz:Korpa:2002} are confirmed almost
quantitatively even though an improved renormalization scheme was applied here.
At zero antikaon momentum $\vec q=0$ a quite broad spectral distribution is obtained with a pronounced
two-peak structure. Including the p- and d-wave contributions has an important effect on the spectral
distribution moving strength from the lower peak to the higher one. Most spectacular are the implications
of switching on scalar and vector mean fields. Significant strength in soft modes at energies around 200-300 MeV
is predicted provided that p-wave interactions are considered. Note also the quite narrow structures seen at zero
antikaon momentum. The latter reflect the presence of
hyperon--nucleon-hole states. As illustrated in Fig. \ref{fig:1b} this effect is sensitive
to the large scalar and vector mean fields, $\Sigma_S = 350$ MeV and $\Sigma_V = 290$ MeV,
suggested by the Dirac phenomenology. Results for vanishing vector mean fields but a finite scalar mean field of
$\Sigma_S = 60 $ MeV, are compared with the full result in Fig. \ref{fig:1b}. In both cases the nucleon energy
at zero momentum and nuclear saturation density is lowered by 60 MeV. Nevertheless, antikaon spectral distributions
arise that are quite distinct at small momenta. We point out that using attractive scalar but repulsive
vector mean fields  implies that the nucleon energy experiences
an attractive shift at small momenta but a repulsive shift at large momenta. Thus the overall impact on the antikaon
spectral function is a subtle average of attractive and repulsive effects.
We note that in a non-relativistic approach a mimic of such effects would require
a strongly energy or momentum dependent nucleon self-energy.

If compared to the latest work by Tolos,
Ramos and Oset \cite{Tolos:Oset:Ramos:2006} significant differences in the shape of the spectral function
are noted. In particular the influence of p-wave scattering is quite dissimilar.
In order to trace the source of such differences we performed computations relying on the
angle-average approximation as used in \cite{Tolos:Oset:Ramos:2006}. The results of those simulations
are included in all figures systematically by additional thin lines. Overall the angle-average approximation
appears quite reliable for the antikaon spectral function. Visible differences are seen only for the case
of zero mean fields but switched on p- and d-waves. To permit a more quantitative comparison with the recent
work by Tolos, Ramos and Oset \cite{Tolos:Oset:Ramos:2006} we provide Figs. \ref{fig:2} and \ref{fig:3}, which give the antikaon self-energy, $\Pi(\omega, \vec q\,)$, as a function of energy and momentum.
The first figure clearly illustrates the dramatic influence of the nucleon mean fields on an antikaon at
zero momentum. At the larger momentum $q= 450$ MeV, as shown in Fig. \ref{fig:3}, the effect of the mean
fields become of minor importance. Depending on his favorite definition the reader may read  an optical potential for
the antikaon off Figs. \ref{fig:2} and \ref{fig:3}. Note that the latter is not a well defined tool once broad spectral
distributions are encountered.

A comparison with the results of \cite{Tolos:Oset:Ramos:2006} reveals striking differences in particular
on the implications of the p-wave channels. Since the angle-averaged approximation appears to be justified
to amazing accuracy for the antikaon self-energy we conclude that the source of such differences must
be due to the use of quite different interactions. In particular one may worry about a possibly strong cutoff
dependence of the results in \cite{Tolos:Oset:Ramos:2006}.

\newpage

\subsection{In-medium properties of the $\Lambda(1405)$ and $\Lambda(1115)$, $\Sigma(1185)$ }

We continue with a discussion of the in-medium properties
of the $J^{P}=\frac{1}{2}^\pm $ hyperons, the $\Lambda(1115)$, $\Sigma(1185)$ and
$\Lambda(1405)$. In order to keep this discussion self-contained we recall the generic
form of the in-medium scattering amplitude (\ref{T-result}). For simplicity
we assume the absence of the $\frac{3}{2}^\pm$ sector, while discussing the
in-medium properties of the  $\frac{1}{2}^\pm$ states. In general the two sectors are
coupled, however, we find that
the $\frac{3}{2}^\pm$ amplitudes have a negligible influence on the
$\frac{1}{2}^\pm$ amplitudes. In a given isospin channel
the  scattering amplitude has the following form
\begin{eqnarray}
&&{\mathcal T}(w_0,\vec w\,) = T^{(p)}_{[11]}(v_0, \vec w\,)\,\left(\frac{1}{2}+
\frac{v_0\,\gamma_0\,-\vec w\cdot \vec \gamma}{2\,\sqrt{v^2}}\right)
\nonumber\\
&& \quad +\,T^{(p)}_{[22]}(v_0, \vec w\,)\,
\left(
\frac{1}{2}-\frac{v_0\,\gamma_0\,-\vec w\cdot \vec \gamma}{2\,\sqrt{v^2}}\right)
\nonumber\\
&& \quad +\,T^{(p)}_{[12]}(v_0, \vec w\,)\left(i\, \frac{|\vec w|}{\sqrt{v^2}}\,\gamma_0
- i\,\frac{v_0\,\vec w }{| \vec w\,|\,\sqrt{v^2}}\cdot \vec \gamma \right)\,,
\nonumber\\
&& T^{(p)}(v_0,\vec w\,)= M^{(p)}(v_0,\vec w\,)\,\Big[ 1- \Delta J^{(p)}(v_0,\vec w\,)\,M^{(p)}(v_0,\vec w\,)\Big]^{-1}\,,
\label{define-mixing-v}
\end{eqnarray}
where $v_0=w_0-\Sigma_V$. The scattering
amplitudes are obtained by the inversion of a 2$\times$2 matrix. The matrix of loop functions
$\Delta J^{(p)}_{[ij]}(v_0,\vec w\,)$
is normalized with respect to free-space, i.e. at zero density the latter vanish identically. The matrix
of source amplitudes $M^{(p)}_{[ij]}(v_0,\vec w\,)$ is fully determined by
free-space s- and p-wave  scattering amplitudes with $J=1/2$ together with the vector mean-field parameter $\Sigma_V$.
The scalar amplitudes, $T^{(p)}_{[ij]}(v_0,\vec w\,)$, reflect the tensor
basis chosen in (\ref{define-mixing-v}). It is useful
to expand the in-medium scattering amplitude in the basis introduced in \cite{Lutz:Korpa:2002}.
The associated amplitudes, $M_{J^P}(w_0,\vec w\,)$, are characterized by well defined angular momentum
and parity. They generalize the amplitudes of (\ref{recoupling-identity}).
It holds
\begin{eqnarray}
&& M_{\frac{1}{2}^\pm}(w_0,\vec w\,) =
\frac{1}{2}\,\Big( \frac{v_0\,w_0-\vec w\,^2}{\sqrt{v_0^2-\vec w\,^2}\,\sqrt{w_0^2-\vec w\,^2}}\mp 1\Big)\,
T^{(p)}_{[11]}(v_0,\vec w\,)
\nonumber\\
&& \qquad  \qquad \quad \;\,\,
-\,\frac{1}{2}\,\Big( \frac{v_0\,w_0-\vec w\,^2}{\sqrt{v_0^2-\vec w\,^2}\,\sqrt{w_0^2-\vec w\,^2}}\pm 1\Big)\,
T^{(p)}_{[22]}(v_0,\vec w\,)
\nonumber\\
&& \qquad \qquad \quad \;\,\,
-\,\frac{1}{2}\, \frac{i\,|\vec w|\,(v_0-w_0)}{\sqrt{v_0^2-\vec w\,^2}\,\sqrt{w_0^2-\vec w\,^2}}\,\Big(
T^{(p)}_{[12]}(v_0,\vec w\,)+T^{(p)}_{[21]}(v_0,\vec w\,)\Big) \,.
\label{w-basis}
\end{eqnarray}

\begin{figure}[t]
\begin{center}
\includegraphics[width=14.5cm,clip=true]{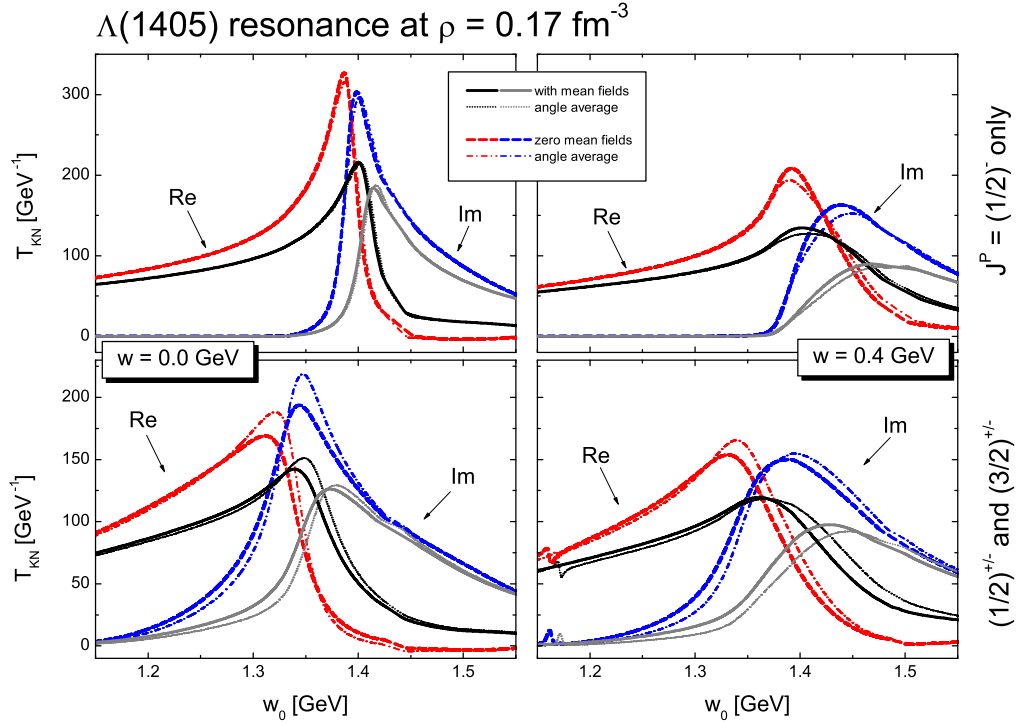}
\end{center}
\caption{$\Lambda(1405)$ mass distribution as a function of energy $w_0$ and momentum $\vec w$
at nuclear saturation density. The results of various approximations are shown. }
\label{fig:4}
\end{figure}

\begin{figure}[t]
\begin{center}
\includegraphics[width=14.5cm,clip=true]{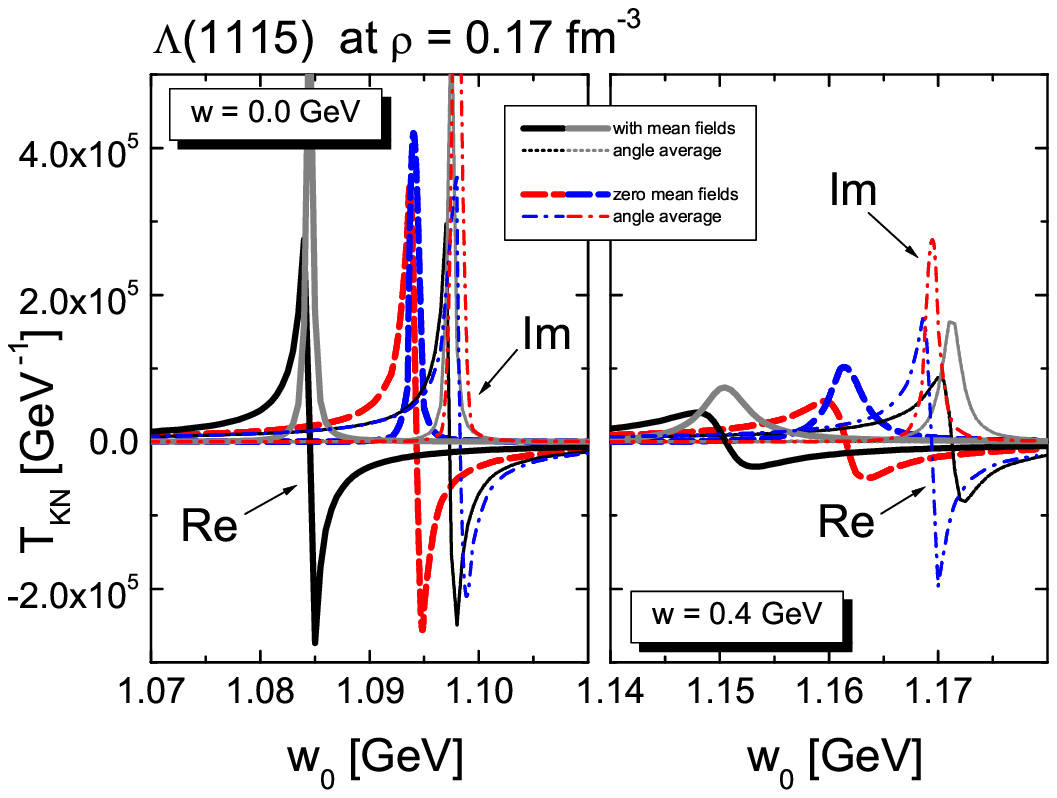}
\end{center}
\caption{$\Lambda(1115)$ mass distribution as a function of energy $w_0$ and momentum $\vec w$
at nuclear saturation density.}
\label{fig:5}
\end{figure}

\begin{figure}[t]
\begin{center}
\includegraphics[width=14.5cm,clip=true]{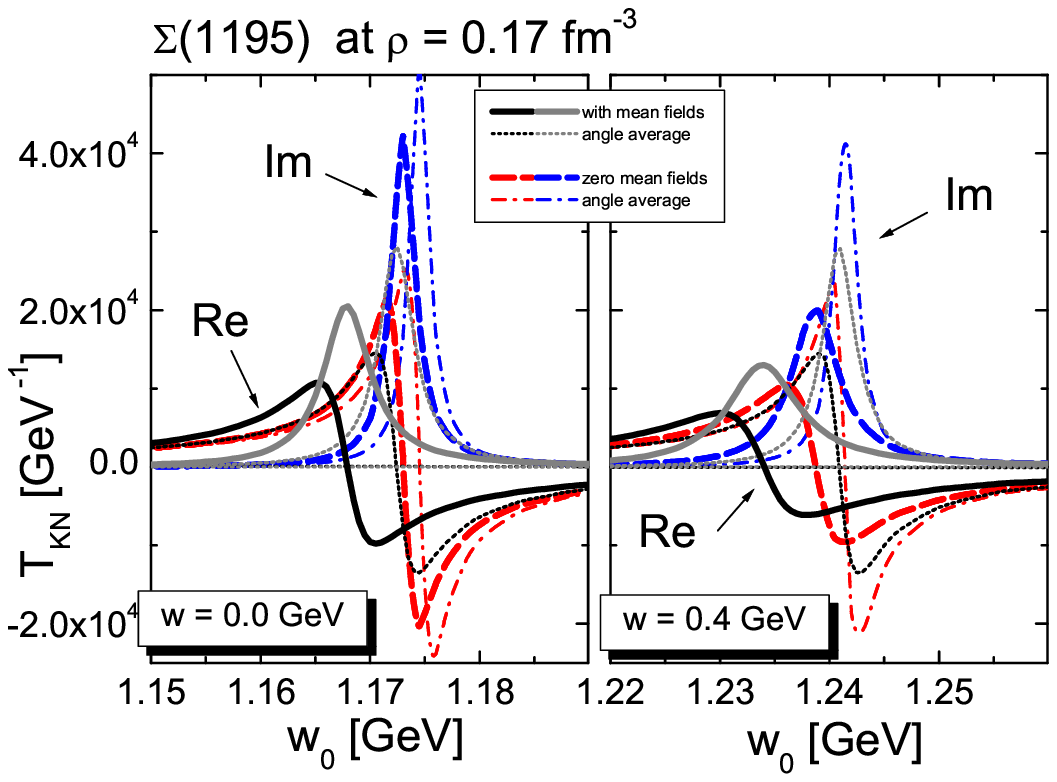}
\end{center}
\caption{$\Sigma(1195)$ mass distribution as a function of energy $w_0$ and momentum $\vec w$
at nuclear saturation density.}
\label{fig:6}
\end{figure}

In Fig. \ref{fig:4} the isospin zero s-wave $\bar K N$ amplitude is shown at saturation density. We confirm the
striking consequence of self-consistency \cite{Lutz:1998,Lutz:Korpa:2002}. Taking into account s-wave
interactions only the resonance is broadened somewhat by the nuclear environment leaving its central
mass unchanged. Switching on scalar and vector mean fields for the nucleon further dissolves the resonance,
as shown in the upper panels of Fig. \ref{fig:4}. Once p-wave interactions are considered, however,
a significant downward shift of about 50 MeV accompanied by further broadening is observed. The mass shift
is reduced in part if the nucleon mean fields are switched on. In all cases the angle-average approximation
appears to work quite reliably. All together the resonance mass is shifted by about 30 MeV only.

We turn to the properties of the hyperon ground states.  For
switched-off or switched-on mean fields a mass shift of about 72
MeV and 80 MeV for the $\Lambda (1115)$ is obtained. The mass
shift is not affected much by the presence of a nucleon mean
field. The angle-average approximation arrives at about 76 MeV and
79 MeV respectively.

We observe that the mass shift of about 72 MeV
derived for switched off mean fields differs significantly from
our previous shift of about 10 MeV only \cite{Lutz:Korpa:2002}.
The difference is a consequence of the improved many-body approach
that eliminates all medium-induced power divergencies. Thus, we
deem our new result, which is manifestly independent of any ad-hoc
cutoff parameter, more reliable.

In order to correct for the overestimate of the mass shift we
implemented an 'intrinsic' repulsive mass shift of 36 MeV. The
appropriate free-space p-wave amplitude is modelled to have a
pole at a mass shifted by that 36 MeV\footnote{The mean-field contribution of the
$\omega$ exchange was estimated in \cite{Kolomeitsev:Voskresensky:2003} to give a repulsive mass
shift of about 54 MeV at saturation density. In a conventional mean field picture that repulsion is
compensated for by a large attractive term implied by a $\sigma$ exchange. However, if the $\sigma$
meson is dominantly a resonant two-pion state, as  suggested by numerous computations based on the
chiral Lagrangian, the role played by the $\sigma$ exchange is highly uncertain. }.
This defines an additional mean field shift of 36 MeV for the $\Lambda (1115)$
state before self consistency is achieved. The implied results for the in-medium
$\Lambda (1115)$ propagator are shown in Fig. \ref{fig:5} for various approximations. The
corresponding antikaon spectral distributions were shown already in Fig. \ref{fig:1}.
All together we arrive at a mass shift of about 30 MeV, which is compatible
with the empirical shift. The figure illustrates the reliability
of the angle-average approximation in the presence of that additional repulsive mean-field.
In particular at larger momenta the error implied by such an approximation
can be as large as 20 MeV in the mass shift.
We confirm our previous result \cite{Lutz:Korpa:2002} that the mass shift is
quite independent of the three momentum but that the in-medium
width is significantly increased as the $\Lambda(1115)$ moves with
respect to the bulk matter.

We turn to Fig. \ref{fig:6}, which shows that
the mass shifts derived for the $\Sigma(1195)$ are small in all considered cases. This confirms our previous results
\cite{Lutz:Korpa:2002}. For switched off mean fields the predicted mass shifts of about 22 MeV is somewhat larger than
our previous shift of about 10 MeV. Here the improvements in the many-body approach are less relevant. Also
scalar and vector means fields do not provide a significant additional mass shift. Like we observed
before \cite{Lutz:Korpa:2002}, the mass and width shifts are quite independent on the three-momentum $\vec w$.
We cannot exclude the need of an 'intrinsic' mass shift like discussed for the $\Lambda(1115)$. However,
since the effective mass of the $\Sigma(1185)$ is not established so far we refrain from doing so.

We note that the present computation can be extended by using
in-medium spectral distributions for the pion and hyperons in the
$\pi \Lambda(1115)$ and $\pi \Sigma(1195)$ channels.

\clearpage
\newpage

\subsection{In medium properties of the $\Sigma(1385)$ and $\Lambda(1520)$}

In the presence of partial-waves with $J= \frac{3}{2}$ further contributions arise in
representation (\ref{define-mixing-v}) of the scattering amplitude. It would be inconvenient to
present figures for all of the 136 amplitudes computed in this work. Like for the spin-one-half system
we focus on those tensor structures present in free-space. The corresponding
amplitudes are readily identified
\begin{eqnarray}
&&M^{(p)}_{\frac{3}{2}^\pm}(w_0,\vec w\,) = {\textstyle{1\over 9}}\,\sum_{i,j=3}^8 C^{\frac{3}{2}^\pm}_{p,[ij]}(v_0,\vec w\,)\,T_{[ij]}^{(p)}(v_0,\vec w\,) \,,
\nonumber\\
&&M^{(q)}_{\frac{3}{2}^\pm}(w_0,\vec w\,) = {\textstyle{1\over 9}}\,\sum_{i,j=1}^2 C^{\frac{3}{2}^\pm}_{q,[ij]}(v_0,\vec w\,)\,T_{[ij]}^{(q)}(v_0,\vec w\,)\,,
\label{recoupling-spin-32}
\end{eqnarray}
where the coefficients $C^{\frac{3}{2}^\pm}_{p,[ij]}$ and $C^{\frac{3}{2}^\pm}_{q,[ij]}$ are
detailed in Appendix B. In the free-space limit the amplitudes of (\ref{recoupling-spin-32}) recover the
amplitude $M_{\frac{3}{2}^\pm}(\sqrt{s}\,)$ of (\ref{recoupling-identity}).

\begin{figure}[b]
\begin{center}
\includegraphics[width=14cm,clip=true]{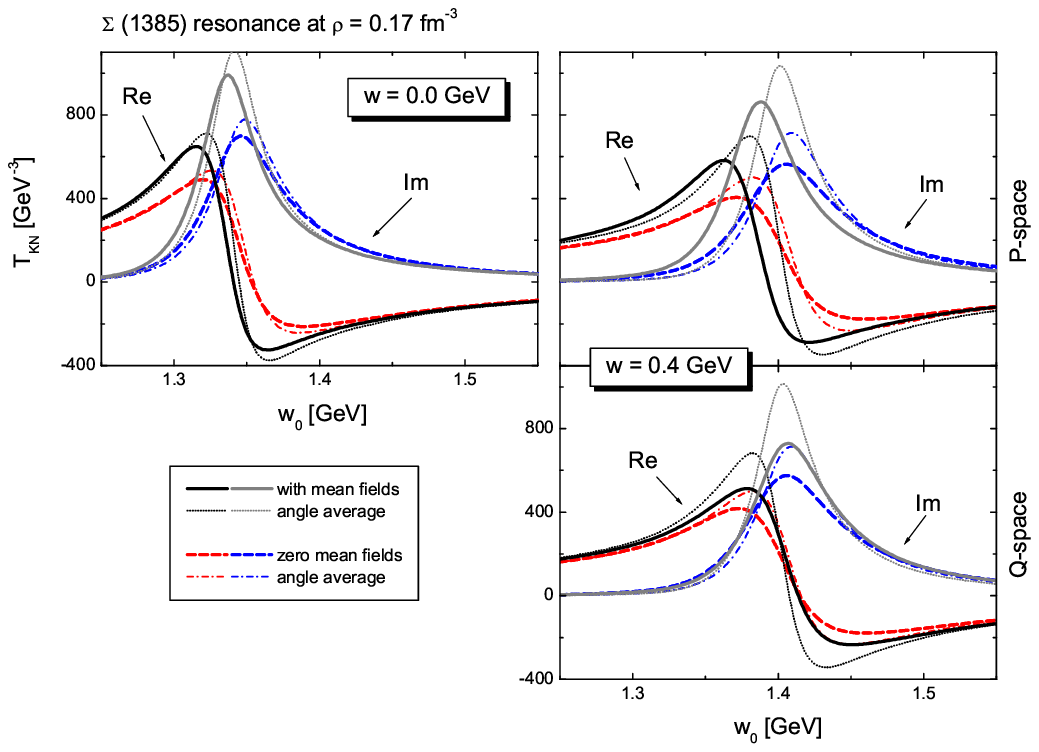}
\end{center}
\caption{$\Sigma(1385)$ mass distribution as a function of energy $w_0$ and momentum $\vec w$
at nuclear saturation density.}
\label{fig:7}
\end{figure}

\begin{figure}[t]
\begin{center}
\includegraphics[width=14cm,clip=true]{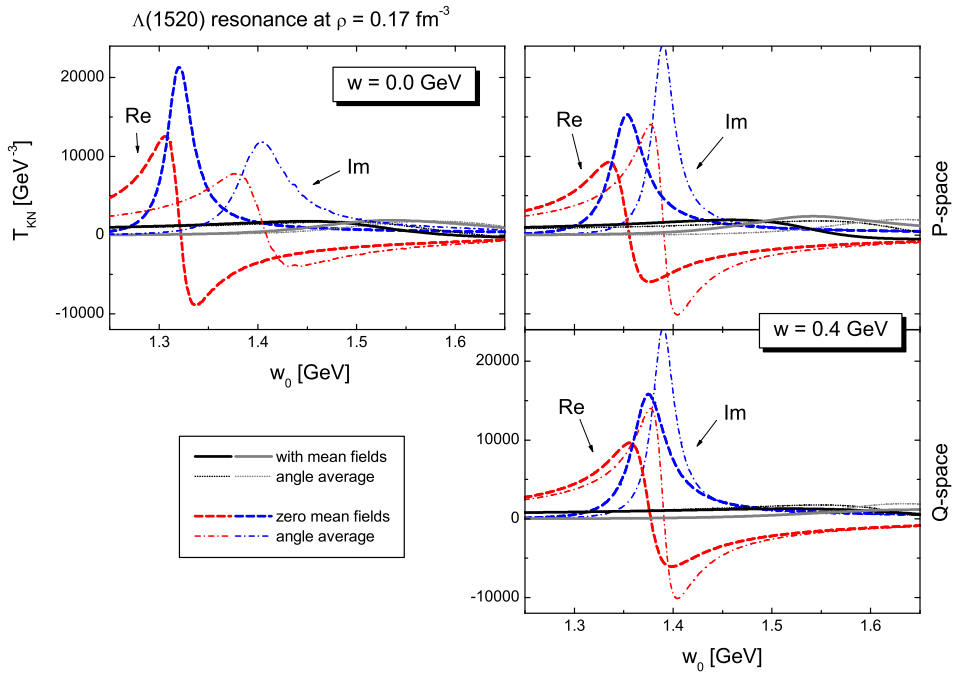}
\end{center}
\caption{$\Lambda(1520)$ mass distribution as a function of energy $w_0$ and momentum $\vec w$
at nuclear saturation density.}
\label{fig:8}
\end{figure}

In Fig. \ref{fig:7} our results for the p-wave $\Sigma(1385)$ resonance are summarized. It is pointed out that
as compared to our previous work \cite{Lutz:Korpa:2002} we obtain a somewhat smaller mass shift. This is a consequence
of the improved renormalization scheme developed in this work. However, once the nucleon mean fields are switched
on we are almost back to our old result. All together we predict an attractive  mass shift of about 40 MeV. This value
is in striking disagreement with the recent result of Tolos, Ramos and Oset \cite{Tolos:Oset:Ramos:2006}, which claim an
attractive shift of 7 MeV only. Since, the angle-average approximation used in \cite{Tolos:Oset:Ramos:2006}
cannot explain this large difference, we conclude that the discrepancy may
be explained by the use of different p-wave amplitude, in particular
at subthreshold energies. One may speculate, that the prescription
devised to treat p-wave effects  or a large dependence on the cutoff parameter
may cause such differences.

In Fig. \ref{fig:8} our results for the p-wave $\Lambda(1520)$ resonance are summarized. Here the effect of
the mean fields are most dramatic. All together the resonance is basically dissolved in nuclear matter already at nuclear
saturation density. It is interesting to observe that for switched off mean fields the resonance receives
an attractive mass shift of about 200 MeV. Here an angle-average approximation would underestimate the shift by
about 80 MeV. This is easily understood: the higher the partial wave the less reliable an angle-average approximation
works.

\clearpage
\newpage

\subsection{Antikaons in strongly compressed nuclear matter}

We conclude the numerical result section by a explorative study of strangeness properties at twice nuclear
saturation density. For such systems it is difficult to establish firm results due to large uncertainties
in the values the scalar and vector mean fields for the nucleon take. Also there is no empirical constraint
on the hyperon ground-state properties at such densities. Effects not considered in this work, like pion dressing
or short-range correlation effects on the hyperon ground states, need to be addressed and controlled. To this
extent the following discussion will be qualitative and should be taken with a grain of salt.

We study four different scenarios. In all four cases the $\Lambda(1115)$ is given an 'intrinsic' repulsive mean field
of 100 MeV at twice saturation density. The latter
was chosen such that the in-medium mass of the $\Lambda(1115)$ after self consistency is pulled down by 25 MeV with
respect to its free-space value for the particular choice $\Sigma_S=\Sigma_V= 500$ MeV.  We deem this as a conservative
estimate. Recall that at saturation density we used a
repulsive 'intrinsic' shift of 36 MeV only, which lead to a mass shift of 30 MeV. The fact that we need such
large repulsive 'intrinsic' mass shifts for the $\Lambda(1115)$ reflects a significant cancellation of repulsive
mean-field type effects and exchange-type effects implied by the strong coupling of the $\Lambda(1115)$ to
the $\bar K N$ channel.

Consider first two large-mean field cases defined by $\Sigma_S=500$ MeV and
\begin{eqnarray}
\Sigma_V = \left\{ \begin{array}{ll}
500 \,{\rm MeV} \\
460 \,{\rm MeV}
\end{array}
\right.
\qquad \qquad {\rm at}\qquad \rho =2\,\rho_0 \,.
\label{mean-fields-2rho0-A}
\end{eqnarray}

Given the nucleon
mean fields (\ref{mean-fields-2rho0-A}) at twice saturation density the
chemical potential is readily estimated
\begin{eqnarray}
\mu = \sqrt{(m_N-\Sigma_S)^2+k_F^2}+ \Sigma_V \simeq
\left\{ \begin{array}{ll}
1055 \,{\rm MeV} \\
1015 \,{\rm MeV}
\end{array}
\right.\quad {\rm at}\quad k_F=340\,{\rm MeV}\,,
\end{eqnarray}
given the mean-field picture.
Since the chemical potential is smaller than the effective $\Lambda(1115)$ mass there is no hyperonization yet
at the considered density.

\begin{figure}[t]
\begin{center}
\includegraphics[width=14cm,clip=true]{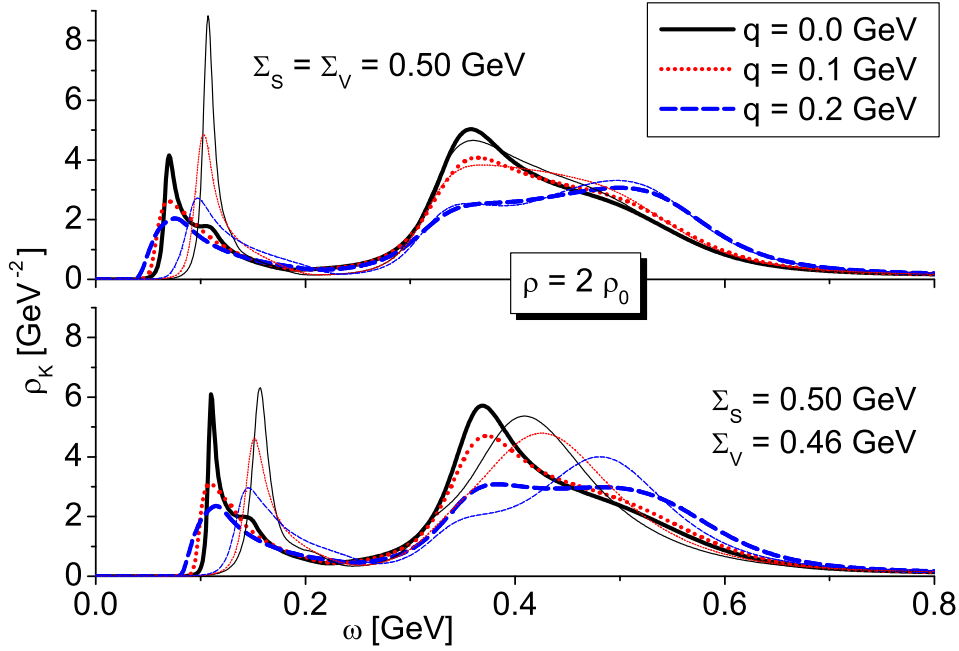}
\end{center}
\caption{Antikaon spectral distribution  as a function of energy $\omega$ and momentum $\vec q$
at twice nuclear saturation density. The thin and thick lines show the results with and without angle-average
approximation. Two large-mean field scenarios are shown. }
\label{fig:9}
\end{figure}

In Fig. \ref{fig:9} the resulting antikaon spectral distributions are shown. A striking effect is revealed. At small
antikaon energies the spectral distributions develops significant strength in a narrow peak at around 70 MeV or 110 MeV
depending on the choice of the mean fields. The
peak remain narrow and pronounced for finite antikaon momenta $0$ MeV $< |\vec q \,| < 200 $ MeV. This is in contrast to
the antikaon spectral distribution at saturation density as shown in Fig. \ref{fig:1b}. The corresponding
structure has very little weight and is dissolved much more quickly as the antikaon starts to move through the matter bulk.
The physical origin of that peak is readily understood: it reflects the coupling of the antikaon to a $\Lambda(1115)$
nucleon-hole state. We emphasize that the soft antikaon mode sits at 70 MeV or 110 MeV, even though the
$\Lambda(1115)$ effective
mass is pulled down by 25 MeV and 23 MeV below its free-space limit at the considered density $2 \,\rho_0$.
In the low-density limit the soft mode has energy
$m_\Lambda-m_N \simeq 175$ MeV, a value significantly larger than the 70 MeV or 110 MeV seen in Fig. \ref{fig:9}.
This illustrates that the $\Lambda(1115)$ nucleon-hole state turns highly collective. The peak positions at
$\vec q=0$ follow quite accurately the difference of the $\Lambda(1115)$ quasi-particle energy and the nucleon
hole-energy at maximum momentum $|\vec p\,| =k_F = 340$ MeV. The complicated antikaon nuclear dynamics appears to
collect maximum strength at the phase-space boundary.

We observe that the angle-average approximation works
less reliably at larger densities. This is illustrated by the thin lines in Fig. \ref{fig:9}, which should be compared to the
thick lines presenting results for the full simulations not relying on any angle-average approximation.
The shifts in the low-mass peaks reflect roughly the different mass shifts for the $\Lambda(1115)$ in the two
approximations. For the two choices of the mean fields the effective $\Lambda(1115)$ masses are found
at 1120 MeV and 1124 MeV relying on the angle-average approximation.

\begin{figure}[t]
\begin{center}
\includegraphics[width=14cm,clip=true]{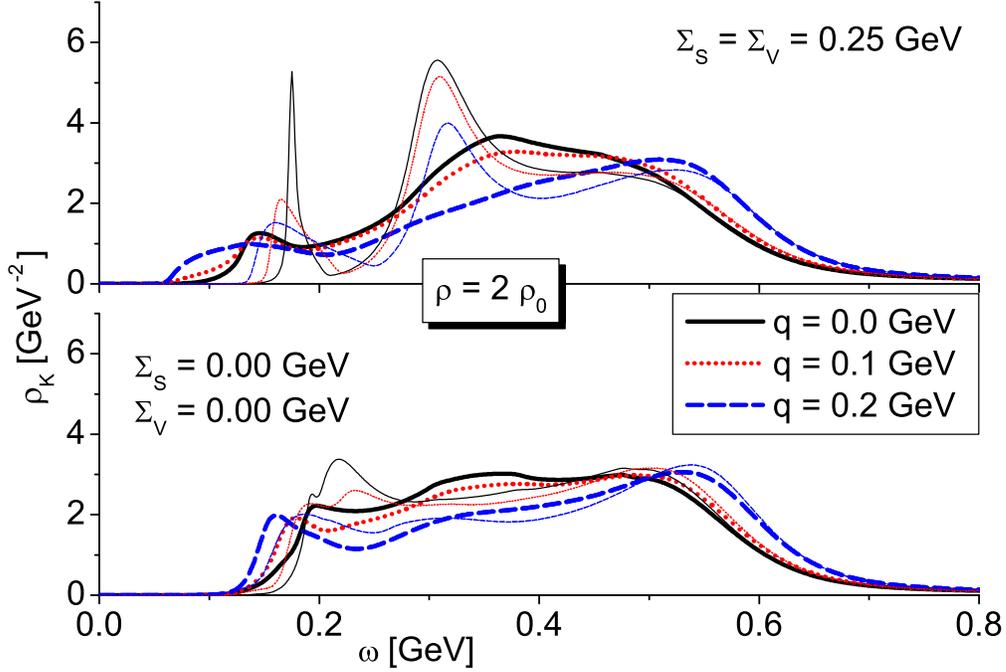}
\end{center}
\caption{Antikaon spectral distribution  as a function of energy $\omega$ and momentum $\vec q$
at twice nuclear saturation density. The thin and thick lines show the results with and without angle-average
approximation. Two small-mean field scenarios are shown.}
\label{fig:10}
\end{figure}

In order to trace the source of the spectacular effects shown in Fig. \ref{fig:9} we performed simulations
at smaller nucleon mean fields as well.
Consider the two small-mean field cases defined by
\begin{eqnarray}
\Sigma_S = \Sigma_V =\left\{ \begin{array}{ll}
250 \,{\rm MeV} \\
0 \,{\rm MeV}
\end{array}
\right. \,,
\qquad
\mu  \simeq
\left\{ \begin{array}{ll}
1018 \,{\rm MeV} \\
\phantom{1}999 \,{\rm MeV}
\end{array}
\right. \qquad {\rm at}\qquad \rho =2\,\rho_0 \,.
\label{mean-fields-2rho0}
\end{eqnarray}
The 'intrinsic' mass shift for the $\Lambda(1115)$ is unchanged as
compared to the large-mean  field cases. In Fig. \ref{fig:10} the resulting antikaon spectral distributions are shown.
The striking low-mass peak structures disappeared in the full simulations of the two small-mean field scenarios. Only
within the angle-average approximation a narrow peak at small mass is seen for the case $\Sigma_S=\Sigma_V=250$ MeV.
For completeness we provide the effective $\Lambda(1115)$ mass underlying the dynamics shown in Fig. \ref{fig:10}.
For the choice $\Sigma_S=\Sigma_V=250$ MeV the effective mass comes at 1078 MeV and 1144 MeV
without and with angle-average approximation. For vanishing mean fields the corresponding values are
1124 MeV and 1126 MeV.

Our findings may have important consequences for the physics of compact stars since antikaon condensation
might occur at moderate densities already (see e.g. \cite{Kolomeitsev:Voskresensky:2003}). Also finite nuclear
systems with strangeness may be affected. Given a finite and compressed nucleus $A$ our results show that the two
states $A\,\Lambda \,N^{-1}$ and $A\,\bar K $ interact strongly with each
other by strangeness-exchange forces if large scalar and vector mean fields for the nucleon are realistic.
The final $A\,\Lambda \,N^{-1}$ state is pulled down to
smaller energies by a significant level-level repulsion of the two states. We thus arrive at
the conclusion that deeply bound and narrow kaonic nuclei may exist as suggested by
Akaishi and Yamazaki  \cite{Akaishi:Yamazaki:2002,Akaishi:Dote:Yamazaki:2005}, however, based on a different mechanism.
For instance an $\alpha$ nucleus offered a strangeness quanta may further shrink in size as a consequence of the
soft antikaon mode as seen in Fig. \ref{fig:9}. Depending on the details the lowest state formed may have
higher nuclear densities than the one of the $\alpha$ particle.

\clearpage
\newpage

\section{Summary}

In this work we generalized the self-consistent and covariant many-body approach \cite{Lutz:Korpa:2002}
for the presence of scalar and vector mean fields of the nucleon. Based on coupled-channel interactions that
were derived from the chiral SU(3) Lagrangian and that were shown to be consistent with low-energy
differential scattering data \cite{Lutz:Kolomeitsev:2002} we
performed numerical simulations of the antikaon and hyperon spectral density in cold nuclear matter.

Without scalar and vector mean fields we confirm our previous results that the consideration of
p-wave scattering in addition to s-wave scattering,  leads to significantly more attraction for
the $\Lambda(1405)$, $\Sigma(1385)$ and $\Lambda(1520)$ resonances. This is ascertained
by an improved renormalization scheme, that avoids any in-medium induced power-divergent
structures as well as the occurrence of kinematical singularities. The latter were regulated in
previous works by ad-hoc cutoffs or form factors. We studied the quality of the angle-average
approximation applied by Oset and collaborators in their many-body approaches to antikaon and hyperon
propagation properties \cite{Ramos:Oset:2000,Tolos:Ramos:Polls:Kuo:2001,Tolos:Oset:Ramos:2006}.
Typically, the angle-average approximation appears sufficient to compute the antikaon spectral function reasonably
well. However, the hyperon ground states and resonances are in part poorly reproduced once the angle-average
approximation is assumed. This is because the p-wave and d-wave phase space of the antikaon-nucleon system is
not always reproduced accurately enough. Most striking is the discrepancy for the d-wave $\Lambda(1520)$ resonance for
which a  difference of about 80 MeV arises.

The effect of incorporating scalar and vector mean fields for the nucleon is important for the
antikaon spectral function that becomes significantly more narrow at small momenta. We emphasize that
it does not suffice to consider a weak scalar mean field. It is crucial to implement both, scalar and vector mean
fields into the self-consistent and covariant many-body approach. Since an attractive scalar but repulsive
vector mean field is used the nucleon energy experiences an attractive shift at small momenta but a repulsive shift
at large momenta. The repulsive effect of the mean fields on the $\Lambda(1405)$-mass shift
as well as on the antikaon spectral function is the result of a subtle average of the two effects.
The mean fields affect the hyperon resonances, with the exception of the $\Lambda(1520)$ resonance, only moderately.
The $\Lambda(1520)$ dissolves almost completely already at saturation density.

Like all previous self-consistent approaches to antikaon and hyperon propagation properties in nuclear matter
we do not substantiate the strong-attraction scenario of Akaishi and Yamazaki. The main antikaon  mode is
pulled down at saturation density by about 50 MeV only. However, at larger nuclear densities we uncovered a novel
phenomenon that could lead to the formation of deeply bound kaonic systems and a novel antikaon-condensation mechanism
in compact stars, if large scalar and vector mean fields for the nucleon are realized in nature. For instance, at twice
nuclear matter densities assuming scalar and vector mean fields of the nucleon degenerate at 500 MeV in magnitude we
obtained a narrow antikaon mode at 70 MeV for antikaon momenta smaller than 200 MeV. The latter reflects a highly
collective $\Lambda(1115)$ nucleon-hole state that is pushed down to lower mass by interaction with the main antikaon
modes. The corresponding effective $\Lambda(1115)$ mass at twice saturation density is 1090 MeV. The precise position
of the soft antikaon mode depends
sensitively on the details of the dynamics. A more quantitative understanding of the proposed mechanism requires
further detailed studies, in particular the role played by short-range correlations.

{\bfseries{Acknowledgments}}
%%%%%%%%%%%%%%%%%%%%%%%%%%%%%%%%%%%%%%%%%%%%%%%%%%%%%%%%%%%%%%%%%%%%%

M.F.M.L. acknowledges useful discussions with E.E. Kolomeitsev, A.~Ramos, F.~Riek, L. Tolos and D.N. Voskresensky.

\newpage

\section*{Appendix A}

We express the projectors in
terms of appropriate building blocks  $P_\pm$, $U_\pm$, $V_\mu$
and $L_\mu, R_\mu$ of the form:
\begin{eqnarray}
&& P_\pm(v) = \frac{1}{2}\left( 1\pm
\frac{\vslash}{\sqrt{v^2}}\right)\, ,\quad U_\pm (v,u)=P_\pm(v)
\,\frac{-i\,\gamma \cdot u}{\sqrt{(v\cdot
u)^2/v^2-1}}\,P_\mp(v)\;,
\nonumber\\
&&V_\mu (v)=\frac{1}{\sqrt{3}}\,\Big( \gamma_\mu
-\frac{\vslash}{v^2}\,v_\mu \Big) \;,\quad X_\mu(v,u)=
\frac{(v\cdot u)\,v_\mu-v^2\,u_\mu}{v^2\,\sqrt{(v \cdot
u)^2/v^2-1}} \;,
\nonumber\\
&&R_\mu (v,u) = +\frac{1}{\sqrt{2}}\,\Big(
U_+(v,u)+U_-(v,u)\Big)\,V_\mu(v)-i\,\sqrt{\frac{3}{2}}\,X_\mu(v,u)
\, , \quad
\nonumber\\
&& L_\mu(v,u) =+\frac{1}{\sqrt{2}}\,V_\mu(v)\, \Big(
U_+(v,u)+U_-(v,u)\Big) -i\,\sqrt{\frac{3}{2}}\,X_\mu(v,u) \;.
\label{def-basic}
\end{eqnarray}
For a compilation of useful properties of the building blocks $P_\pm$, $U_\pm$,
$V_\mu$ and $R_\mu, L_\mu$ we refer to the original work \cite{Lutz:Korpa:2002}.
The q-space projectors are
\begin{eqnarray}
&& Q_{[11]}^{\mu \nu } =\Big( g^{\mu \nu}-\hat v^\mu\,\hat v^\nu
\Big) \,P_+ - V^\mu\,P_-\,V^\nu -L^\mu\,P_+\,R^\nu \;,
\nonumber\\
&& Q_{[22]}^{\mu \nu } =\Big( g^{\mu \nu}-\hat v^\mu\,\hat v^\nu
\Big) \,P_- - V^\mu\,P_+\,V^\nu -L^\mu\,P_-\,R^\nu \;,
\nonumber\\
&& Q_{[12]}^{\mu \nu }  = \Big(g^{\mu \nu}-\hat v^\mu\,\hat v^\nu
\Big)\,U_+ +{\textstyle{1\over 3}}\,V^\mu\,U_-\,V^\nu
\nonumber\\
&&\qquad +{\textstyle{\sqrt{8}\over 3}}\,
\Big( L^\mu\,P_+\,V^\nu +V^\mu\,P_-\,R^\nu \Big) -{\textstyle{1\over 3}}\,L^\mu\,U_+\,R^\nu\;,
\nonumber\\
&&Q_{[21]}^{\mu \nu } = \Big(g^{\mu \nu}-\hat v^\mu\,\hat
v^\nu\Big)\,U_- +{\textstyle{1\over 3}}\,V^\mu\,U_+\,V^\nu
\nonumber\\
&&\qquad +{\textstyle{\sqrt{8}\over 3}}\,
\Big( L^\mu\,P_-\,V^\nu +V^\mu\,P_+\,R^\nu \Big)-{\textstyle{1\over 3}}\,L^\mu\,U_-\,R^\nu\;,
\label{q-space-def}
\end{eqnarray}
where $\hat v_\mu = v_\mu /\sqrt{v^2} $. Using the properties of the building
blocks $P_\pm$, $U_\pm$, $V_\mu$ and $L_\mu, R_\mu$
\cite{Lutz:Korpa:2002} reveals that the objects $Q^{\mu \nu}_{[ij]}$
indeed form a projector algebra.

The p-space projectors have similar transparent representations. Following \cite{Lutz:Korpa:2002}
it is convenient to extend the p-space algebra including objects with one or no Lorentz
index,
\begin{eqnarray}
&&\begin{array}{llll}
P_{[11]} = P_+  \,, & P_{[12]}= U_+ \,, & P_{[21]}=U_-\,, & P_{[22]}=P_- \,, \\
P^\mu_{[31]} = V^\mu \,P_+ \,,  & P^\mu_{[32]} = V^\mu \,U_+ \;, &
\bar P^\mu_{[13]} = P_+\,V^\mu \;,  &  \bar P^\mu_{[23]} = U_-\,V^\mu \;,  \\
P^\mu_{[41]} = V^\mu \,U_- \;,  & P^\mu_{[42]} = V^\mu \,P_- \;, &
\bar P^\mu_{[14]} = U_+\,V^\mu\;,  & \bar P^\mu_{[24]} = P_-\,V^\mu\;,  \\
P^\mu_{[51]} = \hat v^\mu \,P_+ \;,  & P^\mu_{[52]} = \hat v^\mu
\,U_+ \;, &
\bar P^\mu_{[15]} = P_+\,\hat v^\mu \;,  & \bar P^\mu_{[25]} = U_-\,\hat v^\mu \;, \\
P^\mu_{[61]} = \hat v^\mu \,U_- \;,  & P^\mu_{[62]} = \hat v^\mu
\,P_- \;, &
\bar P^\mu_{[16]} = U_+\,\hat v^\mu\;, & \bar P^\mu_{[26]} = P_-\,\hat v^\mu\;, \\
P^\mu_{[71]} =   L^\mu \,P_+ \;,  & P^\mu_{[72]} =   L^\mu \,U_+ \;, &
\bar P^\mu_{[17]} = P_+\,  R^\mu \;,  & \bar P^\mu_{[27]} = U_-\,  R^\mu \;, \\
P^\mu_{[81]} =   L^\mu \,U_- \;,  & P^\mu_{[82]} =   L^\mu \,P_- \;, &
\bar P^\mu_{[18]} = U_+\,  R^\mu\;, & \bar P^\mu_{[28]} = P_-\,  R^\mu\;,
\end{array}
\nonumber\\ \nonumber\\
&& P_{[i\,j]}^{\mu \nu} = P^\mu_{[i1]}\;\bar P^\nu_{[1j]} = P^\mu_{[i2]}\;\bar P^\nu_{[2j]}\,.
\label{p-space-def}
\end{eqnarray}

\newpage

\section*{Appendix B}

We specify the kinematic functions $C_{p,[ij]}^{J^P}(v,u)$ and $C_{q,[ij]}^{J^P}(v,u)$ as introduced in
(\ref{recoupling-identity}). Due to the orthogonality properties of the projectors the latter
are determined by the traces
\begin{eqnarray}
&& C_{p,[ij]}^{\frac{1}{2}^\pm}(v,u)= \Bigg\{
\begin{array}{l}
\frac{1}{2} \,{\tr }\,P_{[ij]}(v,u)\,P^{\frac{1}{2}^\pm}(w) \, \quad \;\;{\rm for} \quad i,j=1,2 \\
0  \qquad \qquad \qquad \qquad \qquad \;\;\,{\rm else}
\end{array} \,,
\nonumber\\
&& C_{p,[ij]}^{\frac{3}{2}^\pm}(v,u) =
\Bigg\{
\begin{array}{l}
\frac{1}{2} \,{\tr }\,P^{\mu \nu}_{[ij]}(v,u)\,P^{\frac{3}{2}^\pm}_{\nu \mu}(w)\, \quad \;\; {\rm for} \quad i,j\neq 1,2 \\
0  \qquad \qquad \qquad \qquad \qquad \;\; \,{\rm else}
\end{array} \,,
\nonumber\\
&& C_{q,[ij]}^{\frac{3}{2}^\pm}(v,u) =
\Bigg\{
\begin{array}{l}
\frac{1}{2} \,{\tr }\,Q^{\mu \nu}_{[ij]}(v,u)\,P^{\frac{3}{2}^\pm}_{\nu \mu}(w)\, \quad \;\; {\rm for} \quad i,j= 1,2 \\
0  \qquad \qquad \qquad \qquad \qquad \;\;\,{\rm else}
\end{array} \,,
\end{eqnarray}
where
\begin{eqnarray}
&&P^{\frac{1}{2}^\pm}(w) = \frac{1}{2}\,\Big(\frac{\wslash }{\sqrt{w^2}} \mp 1 \Big) \,,
\\
&& P^{\frac{3}{2}^\pm}_{\mu \nu}(w) =\frac{3}{2}\,\Big(\frac{\wslash }{\sqrt{w^2}} \pm 1  \Big)\,
\Big\{ \frac{w_\mu\,w_\nu}{w^2}-g_{\mu \nu}+ \frac{1}{3}\,\Big( \gamma_\mu- \frac{\wslash \,w_\mu}{w^2} \Big)\,
\Big( \gamma_\nu- \frac{ \wslash \,w_\nu}{w^2} \Big) \Big\} \,.\nonumber
\end{eqnarray}
The recoupling functions enjoy the symmetry relations
\begin{eqnarray}
&&C^{J^P}_{p,[ij]} =C^{J^P}_{p,[ji]}\,, \qquad \qquad \qquad \;\,C^{J^P}_{q,[ij]} =C^{J^P}_{q, [ji]} \,, \qquad
\nonumber\\
&&C^{J^P}_{p,[5i]} =-\sqrt{3}\,C^{J^P}_{p,[3i]}\,, \qquad \qquad C^{J^P}_{p,[6i]} =+\sqrt{3}\,C^{J^P}_{p,[4i]} \,.
\end{eqnarray}
We derive explicit expressions in the nuclear matter rest frame
\begin{eqnarray}
&&C^{\frac{1}{2}^\pm}_{p,[11]} =-C^{\frac{1}{2}^\mp}_{p,[22]} =
\frac{-1}{3}\,C^{\frac{3}{2}^\mp}_{q,[11]}=\frac{1}{3}\,C^{\frac{3}{2}^\pm}_{q,[22]}
\nonumber\\
&&  \qquad \;\; \,= \frac{1}{2}\,\Big( \frac{v_0\,w_0-\vec w\,^2}{\sqrt{v_0^2-\vec w\,^2}\,\sqrt{w_0^2-\vec w\,^2}}\mp 1\Big)\,,
\nonumber\\
&& C^{\frac{1}{2}^\pm}_{p,[12]} =\frac{-1}{3}\,C^{\frac{3}{2}^\mp}_{q,[12]}=
\frac{1}{2}\, \frac{-i\,|\vec w|\,(v_0-w_0)}{\sqrt{v_0^2-\vec w\,^2}\,\sqrt{w_0^2-\vec w\,^2}} \,,
\end{eqnarray}
and
\begin{eqnarray}
&& C^{\frac{3}{2}^\pm}_{p,[33]}=-C^{\frac{3}{2}^\mp}_{p,[44]}=\frac{1}{3}\,
C^{\frac{3}{2}^\pm}_{p,[55]}=\frac{-1}{3}\,C^{\frac{3}{2}^\mp}_{p,[66]}\,,\qquad \quad
C^{\frac{3}{2}^\pm}_{p,[77]}=-C^{\frac{3}{2}^\mp}_{p,[88]}\,,
\nonumber\\
&&C^{\frac{3}{2}^\pm}_{p,[33]}= \frac{\vec w\,^2\,(v_0-w_0)^2}{3\,\sqrt{v_0^2-\vec w\,^2}^3\sqrt{w_0^2-\vec w\,^2}^3}\,
\Bigg\{ v_0\,w_0-\vec w\,^2 \pm \sqrt{v_0^2-\vec w\,^2}\sqrt{w_0^2-\vec w\,^2}\Bigg\}\,,
\nonumber\\
&&C^{\frac{3}{2}^\pm}_{p,[77]}= \frac{1}{6\,\sqrt{v_0^2-\vec w\,^2}^3\sqrt{w_0^2-\vec w\,^2}^3}\,
\Bigg\{ -9\,(v_0^3\,w_0^3-|\vec w\,|^6)
\nonumber\\
&& \qquad \qquad \qquad \;\;+ |\vec w\,|^2 \,(v_0\,w_0-|\vec w\,|^2)\,\Big[ 5\,(w_0^2+v_0^2)+17\,v_0\,w_0\Big]
\nonumber\\
&& \qquad \qquad \qquad \;\;\pm \,\sqrt{v_0^2-\vec w\,^2}\sqrt{w_0^2-\vec w\,^2}\, \Bigg(-9\,(w_0^2\,v_0^2+|\vec w\,|^4)
\nonumber\\
&& \qquad \qquad \qquad \qquad \qquad \qquad \qquad \;\;+ |\vec w\,|^2\,\Big[v_0^2+w_0^2+16\,v_0\,w_0\Big]\Bigg)\
\Bigg\}\,,
\end{eqnarray}
and
\begin{eqnarray}
&& C^{\frac{3}{2}^\pm}_{p,[47]}=- \,C^{\frac{3}{2}^\mp}_{p,[38]}\,,\qquad \qquad
C^{\frac{3}{2}^\pm}_{p,[48]}=C^{\frac{3}{2}^\mp}_{p,[37]}\,,
\nonumber\\ \nonumber\\
&&C^{\frac{3}{2}^\pm}_{p,[34]}=\frac{+i\,|\vec w\,|^3\,(v_0-w_0)^3}{3\,\sqrt{v_0^2-\vec w\,^2}^3\sqrt{w_0^2-\vec w\,^2}^3}\,,
\nonumber\\
&&C^{\frac{3}{2}^\pm}_{p,[37]}=\frac{-i\,|\vec w\,|\,(v_0-w_0)}{3\,\sqrt{2}\,\sqrt{v_0^2-\vec w\,^2}^3\sqrt{w_0^2-\vec w\,^2}^3}\,
\Bigg\{ 3\,(v_0^2\,w_0^2+|\vec w\,|^4)
\nonumber\\
&& \qquad \qquad \qquad \;\;-|\vec w\,|^2\,\Big[ v_0^2+w_0^2+4\,v_0\,w_0\Big]
\nonumber\\
&& \qquad \qquad \qquad \;\;\pm \,3\,\big(v_0\,w_0-|\vec w\,|^2\big)\,\sqrt{v_0^2-\vec w\,^2}\sqrt{w_0^2-\vec w\,^2}\,
\Bigg\}\,,
\nonumber\\
&&C^{\frac{3}{2}^\pm}_{p,[38]}=\frac{|\vec w\,|^2\,(v_0-w_0)^2}{3\,\sqrt{2}\,\sqrt{v_0^2-\vec w\,^2}^3\sqrt{w_0^2-\vec w\,^2}^3}\,
\Bigg\{ \pm \,\sqrt{v_0^2-|\vec w\,|^2}\,\sqrt{w_0^2-\vec w\,^2}
\nonumber\\
&& \qquad \qquad \qquad \;\;- \,2\,\big(v_0\,w_0-|\vec w\,|^2\big)
\Bigg\}\,,
\nonumber\\
&&C^{\frac{3}{2}^\pm}_{p,[78]}=\frac{+i\,|\vec w\,|\,(v_0-w_0)}{6\,\sqrt{v_0^2-\vec w\,^2}^3\sqrt{w_0^2-\vec w\,^2}^3}\,
\Bigg\{3\,(v_0^2\,w_0^2 +| \vec w\,|^4)
\nonumber\\
&& \qquad \qquad \qquad \;\;+
| \vec w\,|^2\,\Big[ v_0^2+w_0^2-8\,v_0\,w_0\Big] \Bigg\}\,.
\end{eqnarray}

\newpage

\section*{Appendix C}

We recall the form of the invariant functions
$c^{(p,q)}_{[ij]}(q;w,u)$:
\begin{eqnarray}
&&c_{[11]}^{(q)}={\textstyle{1\over 2}}\,E_+\,\Big(E_+\,E_- +(X \cdot q)^2
\Big)\,,\quad c_{[11]}^{(p)}=E_+\,,\quad
\nonumber\\
&&c_{[12]}^{(q)}= -{\textstyle{i\over 2}}\,(X \cdot q)\,\Big( E_+\,E_- +(X
\cdot q)^2\Big)\,,\quad c_{[12]}^{(p)}=-i\,(X \cdot q)\,,
\nonumber\\
&& c_{[22]}^{(q)}={\textstyle{1\over 2}}\,E_-\,\Big(E_+\,E_- +(X \cdot q)^2
\Big) \,, \quad c_{[22]}^{(p)}=E_-\,,
\nonumber\\
&&c_{[13]}^{(p)}=c_{[24]}^{(p)}=
-{\textstyle{1\over \sqrt{3}}}\,E_+\,E_-\,,\quad
c_{[25]}^{(p)}=c_{[16]}^{(p)}=-i\,(\hat w \cdot q)\,(X \cdot q)\,,
\nonumber\\
&&c_{[17]}^{(p)}= -i\,\sqrt{{\textstyle{2\over 3}}}\,E_+ \,(X \cdot
q)\,,\quad c_{[15]}^{(p)}= (\hat w \cdot q)\,E_+\,,\quad
c_{[14]}^{(p)}= {\textstyle{i\over \sqrt{3}}}\,E_+\,(X \cdot q) \,,
\nonumber\\
&&c_{[28]}^{(p)}= -i\,\sqrt{{\textstyle{2\over 3}}}\,E_- \,(X \cdot q)\,,
\quad c_{[26]}^{(p)}= (\hat w \cdot q)\,E_-\,,\quad
c_{[23]}^{(p)}= {\textstyle{i\over \sqrt{3}}}\,E_-\,(X \cdot q)\,,
\nonumber\\
&&c_{[27]}^{(p)}=c_{[18]}^{(p)}=
-\sqrt{{\textstyle{3\over 2}}}\,\Big({\textstyle{1\over 3}}\,E_+ \,E_- +(X \cdot q)^2
\Big)\,,
\end{eqnarray}
and
\begin{eqnarray}
&&c_{[33]}^{(p)}= {\textstyle{1\over 3}}\,E^2_-\,E_+\,,\quad
c_{[44]}^{(p)}={\textstyle{1\over 3}}\,E_+^2\,E_- \;,
\nonumber\\
&&c_{[55]}^{(p)}= E_+\,(\hat w\cdot q)^2\,,\quad
c_{[77]}^{(p)}={\textstyle{1\over 2}}\,E_+\,\Big({\textstyle{1\over 3}}\,E_+\,E_- -\big(
X\cdot q\big)^2 \Big)\;,
\nonumber\\
&&c_{[66]}^{(p)}= E_-\,(\hat w\cdot q)^2\;,\quad
c_{[88]}^{(p)}={\textstyle{1\over 2}}\,E_-\,\Big({\textstyle{1\over 3}}\,E_+\,E_- -\big(
X\cdot q\big)^2 \Big)\,,
\nonumber\\
&&c_{[35]}^{(p)}=c_{[46]}^{(p)}= -{\textstyle{1\over \sqrt{3}}}\,(\hat w\cdot
q)\,E_+\,E_- \,,\quad c_{[57]}^{(p)}= -i\,\sqrt{{\textstyle{2\over 3}}}\,(X
\cdot q) \,(\hat w\cdot q)\,E_+\,,
\nonumber\\
&&c_{[37]}^{(p)}=c_{[48]}^{(p)}= i\,\frac{\sqrt{2}}{3}\,(X\cdot
q)\,E_+\,E_- \,,\quad c_{[68]}^{(p)}= -i\,\sqrt{{\textstyle{2\over 3}}}\,(X
\cdot q) \,(\hat w\cdot q)\,E_- \;,
\nonumber\\
&&c_{[34]}^{(p)}=  -{\textstyle{i\over 3}}\,(X \cdot q)\,E_+\,E_-\,,\quad
c_{[56]}^{(p)}= -i\,(\hat w \cdot q)^2 \, (X \cdot q) \;,\;\;\;\;
\nonumber\\
&&c_{[78]}^{(p)}= i\,\big( X\cdot q\big)\left( {\textstyle{3\over 2}}\, \big(
X\cdot q\big)^2 +{\textstyle{5\over 6}}\,E_+\,E_- \right)\,,
\nonumber\\
&&c_{[36]}^{(p)}= {\textstyle{i\over \sqrt{3}}}\,(\hat w \cdot q)\,E_- \,(X
\cdot q) \,,\quad c_{[38]}^{(p)}={\textstyle{1\over \sqrt{2}}}\,E_- \,
\Big({\textstyle{1\over 3}}\,E_+\,E_- +(X \cdot q)^2\Big) \;,
\nonumber\\
&&c_{[45]}^{(p)}= {\textstyle{i\over \sqrt{3}}}\,(\hat w \cdot q)\,E_+ \,(X
\cdot q) \;,\quad c_{[47]}^{(p)}={\textstyle{1\over \sqrt{2}}}\,E_+ \,
\Big({\textstyle{1\over 3}}\,E_+\,E_- +(X \cdot q)^2\Big) \;,\quad
\nonumber\\
&&c_{[58]}^{(p)}=c_{[67]}^{(p)}=-\sqrt{{\textstyle{3\over 2}}}\,(\hat w \cdot
q) \,\Big( {\textstyle{1\over 3}}\,E_+\,E_- +(X \cdot q)^2\Big)\,, \label{}
\end{eqnarray}
where
\begin{eqnarray}
&& X_\mu =
\frac{(w\cdot u)\,w_\mu-w^2\,u_\mu}{w^2\,\sqrt{(w \cdot
u)^2/w^2-1}}\,, \qquad \hat w_\mu= \frac{w_\mu}{\sqrt{w^2}}\,,
\nonumber\\
&& E_\pm \equiv m_N\pm (\sqrt{w^2}-q\cdot \hat w )
\;, \qquad E_+\,E_- =q^2-(q\cdot \hat w)^2 \;.
\end{eqnarray}

\newpage

\section*{Appendix D}

We specify first the  loop matrix $J_{[ij]}(v,u)=J_{[ji]}(v,u)$
introduced in (\ref{j-exp}).  They are expressed in terms of the 13 master kernels $J_i(v,u)$ defined
in (\ref{def-Kis}). It holds:
\begin{eqnarray}
&&J^{(q)}_{[11]}=\bar m_N\,J_3+J_7\;, \qquad J^{(q)}_{[22]}=\bar m_N\,J_3-J_7
\;, \qquad J^{(q)}_{[12]}=-i\,J_8\,,
\nonumber\\
&&J^{(p)}_{[11]} = \bar m_N\,J_{0}+J_{1} \;, \qquad J^{(p)}_{[22]} =
\bar m_N\,J_{0}-J_{1} \;,\qquad J^{(p)}_{[12]} = -i\,J_{2}\,,
\nonumber\\
&&J^{(p)}_{[13]}=J^{(p)}_{[24]}= {\textstyle{-1 \over \sqrt{3}}}
\,\Big( 2\,J_3-J_5 \Big)\,,\qquad
J^{(p)}_{[16]}=J^{(p)}_{[25]}=+i\left(
J_{6}-\sqrt{v^{2}}J_{2}\right) ,
\nonumber\\
&&J^{(p)}_{[15]}=+(\sqrt{v^{2}}-\bar m_N)\,J_{1}-J_{4}+\bar m_N\,\sqrt{v^{2}}\,J_{0}\;, \qquad
\nonumber\\
&&
J^{(p)}_{[26]}=-(\sqrt{v^{2}}+\bar m_N)\,J_{1}+J_{4}+\bar m_N\,\sqrt{v^{2}}\,J_{0}\;,
\nonumber\\
&&J^{(p)}_{[17]}= -i\,\sqrt{{\textstyle{2\over 3}}}\,\big( \bar m_N
\,J_2+J_6\big)\;,\qquad J^{(p)}_{[28]}=
-i\,\sqrt{{\textstyle{2\over 3}}}\,\big( \bar m_N \,J_2-J_6\big)\,,
\nonumber\\
&&J^{(p)}_{[14]}= +{\textstyle{i \over \sqrt{3}}}(\bar m_N \,J_2+J_6)\;,\qquad \quad
J^{(p)}_{[23]}= +{\textstyle{i \over \sqrt{3}}}(\bar m_N\, J_2-J_6)\;,
\nonumber\\
&&J^{(p)}_{[18]}=J^{(p)}_{[27]}= -\sqrt{{\textstyle{2\over 3}}}\,\,\Big(
J_3+J_5 \Big)\,,\;\;\; \label{decomp-a}
\end{eqnarray}
and
\begin{eqnarray}
&& J^{(p)}_{[33]}={\textstyle{1\over 3}}\,\Big(\bar m_N\,\big(2\,J_3-J_5 \big)
+J_{12}-2\,J_7 \Big)\,,
\nonumber\\
&& J^{(p)}_{[44]}={\textstyle{1\over 3}}\,\Big(\bar m_N\,\big(2\,J_3-J_5 \big)
-J_{12}+2\,J_7 \Big)\,,
\nonumber\\
&&
J^{(p)}_{[55]}=\big(\bar m_N-2\,\sqrt{v^{2}}\big)\,J_{4}+J_{9}+\bar m_N\,v^{2}\,J_{0}+\big(v^{2}-2\,\bar m_N\,\sqrt{v^{2}}\big)\,J_{1}\,,
\nonumber\\
&&
J^{(p)}_{[66]}=(\bar m_N+2\,\sqrt{v^{2}})\,J_{4}-J_{9}+\bar m_N\,v^{2}\,J_{0}-(v^{2}+2\,\bar m_N\,\sqrt{v^{2}})\,J_{1}\,,
\nonumber\\
&& J^{(p)}_{[77]}={\textstyle{1\over 3}}\,\Big(\bar m_N\,\big(J_3-2\,J_5 \big)
+J_{7}-2\,J_{12} \Big)\,,
\nonumber\\
&& J^{(p)}_{[88]}={\textstyle{1\over 3}}\,\Big(\bar m_N\,\big(J_3-2\,J_5 \big)
-J_{7}+2\,J_{12} \Big)\,,
\nonumber\\
&& J^{(p)}_{[35]}= J^{(p)}_{[46]}={\textstyle{1 \over \sqrt{3}}}\left(
2\,J_{7}-J_{12}-\sqrt{v^{2}}\,\big(2\,J_{3}-J_{5}\big)\right) \,,
\nonumber\\
&&J^{(p)}_{[37]}=
J^{(p)}_{[48]}=i\,{\textstyle{\sqrt{2}\over 3}}\,\Big(2\,J_8-J_{11}\Big)\,,
\nonumber\\
&&J^{(p)}_{[57]}=i\,\sqrt{{\textstyle{2\over 3}}}\left( (\bar m_N
-\sqrt{v^{2}})\,J_{6}+
J_{10}-\bar m_N\sqrt{v^{2}}\,J_{2}\right) \,,\nonumber\\
&&J^{(p)}_{[68]}=i\,\sqrt{{\textstyle{2\over 3}}}\left( (\bar m_N
+\sqrt{v^{2}})J_{6}-J_{10}-\bar m_N\sqrt{v^{2}}\,J_{2}\right)\,,
\nonumber\\
&&J^{(p)}_{[34]}=-{\textstyle{i\over 3}}\,\Big(2\,J_8-J_{11} \Big)\,,\quad
J^{(p)}_{[78]}={\textstyle{i\over 3}}\,\Big(5\,J_8+2\,J_{11} \Big)\,,
\nonumber\\
&&J^{(p)}_{[36]}=-{\textstyle{i \over \sqrt{3}}}\left( (\bar m_N +\sqrt{v^{2}})\,J_{6}-J_{10}-\bar m_N\sqrt{v^{2}}\,J_{2}\right)\,,\nonumber\\
&&J^{(p)}_{[45]}=-{\textstyle{i \over \sqrt{3}}}\left( (\bar m_N
-\sqrt{v^{2}})J_{6}+J_{10}-\bar m_N\sqrt{v^{2}}\,J_{2}\right)\,,
\nonumber\\
&&J^{(p)}_{[38]}={\textstyle{\sqrt{2}\over 3}}\,\Big( \bar m_N\,\big(J_3+J_5
\big)-J_7-J_{12}\Big)\,,
\nonumber\\
&&J^{(p)}_{[47]}={\textstyle{\sqrt{2}\over 3}}\,\Big( \bar m_N\,\big(J_3+J_5
\big)+J_7+J_{12}\Big)\,,
\nonumber\\
&&J^{(p)}_{[58]}=J^{(p)}_{[67]}=\sqrt{{\textstyle{2\over 3}}}\left(
J_{7}+J_{12}-\sqrt{v^{2}}\,(J_{3}+J_{5})\right) \,,\nonumber\\
&&J^{(p)}_{[56]}=-i\left(
J_{10}-2\,\sqrt{v^{2}}\,J_{6}+v^{2}\,J_{2}\right)\,. \label{decomp-b}
\end{eqnarray}
It is understood that
the scalar nucleon mass $\bar m_N = m_N-\Sigma_S$ as specified
in (\ref{def-mean-field}) is used in (\ref{decomp-a},\ref{decomp-b}).

\newpage

\section*{Appendix E}

We provide the renormalized form of the integrals, $J_i(v,u)$, as defined in (\ref{def-Kis})).
It is left to specify the form of the integral kernels $K_i^R(l,\bar v; v,u)$
as introduced in (\ref{def-loop-J_n}). We introduce
\begin{eqnarray}
&& K_0^R = \frac{v^2}{\bar v^2} \,,
\nonumber\\
&& K^R_1 = \left( \frac{\sqrt{v^2}}{2}+ \frac{(\bar l \cdot \bar v)}{\sqrt{v^2}} \right)\, \frac{v^2}{\bar v^2}\,,
\nonumber\\
&& K_2^R = - \frac{(v \cdot u)}{\sqrt{(v \cdot u)^2-v^2}}
\,\frac{(\bar l \cdot \bar v)}{\sqrt{v^2}}  \,\frac{v^2}{\bar v^2}
\, +\frac{\sqrt{v^2}\,(\bar l \cdot u)}{\sqrt{(v \cdot u)^2-v^2}}\,,
\nonumber\\
&& K_3^R =\frac{1}{2}\,K_5^R
-\frac{1}{2}\,\frac{v^2}{\bar v^2} \left( \frac{(\bar l \cdot \bar v)^2}{v^2} - \bar l\,^2 - \frac{\bar v^2-v^2}{4}\right)
\,,
\nonumber\\
&& K_4^R = \left(\frac{\sqrt{v^2}}{2}+ \frac{(\bar l \cdot \bar v)}{\sqrt{v^2}}\right)^2\, \frac{v^2}{\bar v^2}
\,,
\nonumber\\
&& K_5^R= \frac{1}{(v \cdot u)^2-v^2}\,\Bigg\{
v^2\,(\bar l \cdot  u)^2
- 2\,(v \cdot u)\,(\bar l \cdot \bar v)\,(\bar l \cdot u)
\nonumber\\
&& \qquad +\,
(v \cdot u)^2\,\frac{(\bar l \cdot \bar v)^2}{\bar v^2}
\Bigg\}
-\frac{\bar v^2-v^2}{12\,\bar v^2}\,v^2
\,,
\nonumber\\
&& K_6^R  =\frac{\sqrt{v^2}}{2}\,K_2^R +
\frac{(\bar l \cdot \bar v)\,(\bar l \cdot u)}{\sqrt{(v \cdot u)^2-v^2}}
- \frac{(v \cdot u)}{\sqrt{(v \cdot u)^2-v^2}}
\,\frac{(\bar l \cdot \bar v)^2}{\bar v^2}\,,
\nonumber\\
&& K_7^R =\frac{1}{2}\,K_{12}^R
+ \frac{1}{2}\,\frac{v^2}{\bar v^2}\left( \bar l\,^2 + \frac{\bar v^2-v^2}{4} - \frac{(\bar l \cdot \bar v)^2}{ v^2}\right)\,\left( \frac{\sqrt{v^2}}{2}+ \frac{(\bar l \cdot \bar v)}{\sqrt{v^2}} \right)
\,,
\nonumber\\
&&K_8^R =\frac{1}{2}\,K_{11}^R
+ \frac{1}{2}\,\left( \bar l\,^2 + \frac{\bar v^2-v^2}{4} - \frac{(\bar l \cdot \bar v)^2}{v^2}\right)\left(- \frac{(v \cdot u)}{\sqrt{(v \cdot u)^2-v^2}}
\,\frac{(\bar l \cdot \bar v)}{\sqrt{v^2}} \frac{v^2}{\bar v^2}
\nonumber\right.\\
&&\left. \qquad +\frac{\sqrt{v^2}\,(\bar l \cdot u)}{\sqrt{(v \cdot u)^2-v^2}}\right)
\,,
\nonumber\\
&& K_9^R =  \left(\frac{\sqrt{v^2}}{2}+ \frac{(\bar l \cdot \bar v)}{\sqrt{v^2}}\right)^3 \frac{v^2}{\bar v^2}
\,,
\nonumber\\
&& K_{10}^R = -\frac{v^2}{4}\,\frac{v^2}{\bar v^2}\left(1+ 2\,\frac{(\bar l\cdot \bar v)}{ v^2}
+4\,\frac{(\bar l\cdot \bar v)^2}{ v^2\, v^2}\right)\,\frac{(v \cdot u)}{\sqrt{(v \cdot u)^2-v^2}}
\,\frac{(\bar l \cdot \bar v)}{\sqrt{v^2}}\nonumber\\
&&\qquad+\frac{v^2}{4}\left(1+ 2\,\frac{(\bar l\cdot \bar v)}{ v^2}+4\,\frac{(\bar l\cdot \bar v)^2}{ v^2\, v^2}\right)\,\frac{\sqrt{v^2}\,(\bar l \cdot u)}{\sqrt{(v \cdot u)^2-v^2}}
\nonumber\\
&&\qquad
+ \frac{1}{2}\, \frac{\sqrt{v^2}\,(v \cdot u)}{\sqrt{(v \cdot u)^2-v^2}}\,\left( \frac{(\bar l \cdot u)}{( v \cdot u)}
- \frac{(\bar l \cdot \bar v)}{\bar v^2} \right) \,(\bar l \cdot \bar v)
\,,
\nonumber\\
&& K_{11}^R = \frac{\sqrt{v^2}}{[(v \cdot u)^2-v^2]^{3/2}}\,\Bigg\{
v^2\,(\bar l \cdot  u)^3
-\, 3\,(v \cdot u)\,(\bar l \cdot u)^2\,(\bar l \cdot \bar v)\,\frac{v^2}{\bar v^2}
\nonumber\\
&& \qquad  +\, 3\,(v \cdot u)^2\,(\bar l \cdot u)\,\frac{(\bar l \cdot \bar v)^2}{ v^2}
-\,\frac{(v \cdot u)^3}{v^2}\,\frac{(\bar l \cdot \bar v)^3}{\bar v^2}
\Bigg\}
\,,
\nonumber\\
&& K_{12}^R = \frac{1}{(v \cdot u)^2-v^2}\,\Bigg\{
\left( \frac{\sqrt{v^2}}{2}+ \frac{(\bar l \cdot \bar v)}{\sqrt{v^2}}\right)\,\frac{v^2}{\bar v^2}\,v^2\, (\bar l \cdot u)^2
\nonumber\\
&& \qquad -2\, \left( \frac{\sqrt{v^2}}{2}+
\sqrt{v^2}\,\frac{(\bar l \cdot \bar v)}{ v^2}\right)(v\cdot u)\,(\bar l \cdot u)\,(\bar l \cdot \bar v)
\nonumber\\
&& \qquad  + \left( \frac{\sqrt{v^2}}{2}+ \frac{(\bar l \cdot \bar v)}{\sqrt{v^2}}
\right)
(v \cdot u)^2\,\frac{(\bar l \cdot \bar v)^2}{\bar v^2}
\Bigg\}
\,,
\label{KR_i-definitions}
\end{eqnarray}
with
\begin{eqnarray}
&& \bar l_\mu = l_\mu - \bar v_\mu /2 \,, \qquad  \bar v^2 = (\bar
v \cdot u)^2 - (v \cdot u)^2 +v^2\,.
\end{eqnarray}

\newpage

\section*{Appendix F}

The subtraction terms, $J_i^C(v,u)$, of (\ref{disp-integral}) are written
in terms of the coefficients, $C_{a,n}^{ijk}$, introduced in (\ref{def-loop-C_n}). It is derived:
\begin{eqnarray}
&&J^C_{0}=(v \cdot u)\,C_{0,1}^{000} \,, \qquad \qquad \qquad \qquad
J^C_{1}=\frac{1}{2}\,\frac{v \cdot u}{\sqrt{v^2}}\left(v^2\,C_{0,1}^{000} + 2\,C_{0,1}^{100}\right)\,,
\nonumber\\
&&J^C_{2}=-
\frac{v \cdot u}{\sqrt{(v \cdot u)^2 - v^2}}\,\frac{v \cdot u}{\sqrt{v^2}}\,C_{0,1}^{100}\,,
\nonumber\\
&&J^C_{3}=\frac{1}{2}\,J^C_{5}+2\,(v \cdot u)\,C_{0,2}^{200} + 2\,C_{+1,2}^{200} \,,
\nonumber\\
&&J^C_{4}=\frac{1}{4}\,(v \cdot u)\left(v^2\,C_{0,1}^{000} + 4\,C_{0,1}^{100}\right)\,,
\qquad \qquad
J^C_{5}=\frac{2\,(v \cdot u)\,}{(v \cdot u)^2 - v^2}\,C_{-1,0}^{110}\,,
\nonumber \\
&&J^C_{6}=\frac{1}{2}\,\sqrt{v^2}\,J^C_{2}-\frac{1}{\sqrt{(v \cdot u)^2 - v^2}}\,C_{-1,0}^{110}\,,
\nonumber\\
&&J^C_{7}=\frac{1}{2}\,J^C_{12} + \frac{1}{2\,\sqrt{v^2}}\,C_{+1,2}^{300}
-\frac{1}{16}\,\sqrt{v^2}\,v^2\,\left(C_{+1,1}^{000} + 4\,C_{+1,2}^{001} - 16\,C_{+1,3}^{200}\right)
\nonumber\\
&& \quad -\frac{1}{16}\,\sqrt{v^2}\,(v\cdot u)\left(v^2\,C_{0,1}^{000} + 2\,C_{0,1}^{100} - 4\,C_{0,2}^{200} +
        16\,C_{0,3}^{300}\right)
\nonumber\\
&& \quad -\frac{1}{8}\,\sqrt{v^2} \left(C_{+1,1}^{100} +
      4\,C_{+1,2}^{101} - 8\,C_{+1,2}^{200} - 8\,C_{+1,3}^{300}\right)\,,
\nonumber\\
&&J^C_{8}=\frac{1}{2}\,J^C_{11}-\frac{1}{24\,\,\sqrt{v^2}\sqrt{(v \cdot u)^2 - v^2}}\left[v^2\,\left(3\,C_{+1,0}^{010} +
    12\,C_{+1,1}^{011} + 8\,C_{+1,2}^{210}\right) \right.
    \nonumber\\
&&\quad\left.-12\,C_{+1,1}^{210} -   (v \cdot u)^2\,\left(3\,v^2\,C_{0,1}^{100} + 8\,C_{0,2}^{300}\right)
+12\,(v \cdot u)\,\left( C_{+1,2}^{300}-C_{0,1}^{210} \right)
\nonumber\right.\\
&&\quad\left. +
    v^2\,(v \cdot u)\left(3\,C_{0,0}^{010} + 12\,C_{0,1}^{011} - 3\,C_{+1,1}^{100}
   - 12\,C_{+1,2}^{101} + 8\,C_{0,2}^{210} - 8\,C_{+1,3}^{300}\right)\right] \,,
\nonumber\\
&&J^C_{9}=-\frac{1}{8\,\sqrt{v^2}}\left[-(v \cdot u)\,v^2\,(v^2\,C_{0,1}^{000} + 6\,C_{0,1}^{100}) +
  16\,(v \cdot u)\,C_{0,2}^{300} + 8\,C_{+1,2}^{300}\right] \,,
 \nonumber\\
&&J^C_{10}=-\frac{1}{12\,\sqrt{v^2}\,\sqrt{(v \cdot u)^2 - v^2}}\left[
v^2\,\left(6\,C_{-1,0}^{110} + 6\,C_{+1,1}^{110}+16\,C_{+1,2}^{210}\right)
\right.\nonumber\\
&&\left.\quad + 12\,C_{+1,1}^{210} + (v \cdot u)^2\,\left(3\,v^2\,C_{0,1}^{100} -
    16\,C_{0,2}^{300}\right)
+12\,(v \cdot u)\,\left(C_{0,1}^{210} - C_{+1,2}^{300}\right)
     \right.\nonumber\\
&&\left.\quad + 2\,(v \cdot u)\, v^2\,\left(3\,C_{0,1}^{110} - 3\,C_{+1,2}^{200} + 8\,C_{0,2}^{210} -
      8\,C_{+1,3}^{300}\right)\right] \,,
\nonumber\\
&&J^C_{11}=-\frac{1}{\sqrt{v^2}\,((v \cdot u)^2 - v^2)^{3/2}}\left[-2\,(v \cdot u)\,(v ^2)^2\,C_{0,2}^{210} + 2\,(v ^2)^2\,C_{+1,2}^{210} \nonumber\right.\\
&&\left.\quad+ (v \cdot u)^2\,\left(3\,v^2\,C_{0,1}^{120} + 3\,C_{+1,1}^{210} -
    2\,v^2\,C_{+1,2}^{210}\right) \right.\nonumber\\
&&\left.\quad+ (v \cdot u)^3\,\left(3\,C_{0,1}^{210} +
    2\,v^2\,C_{0,2}^{210} - C_{+1,2}^{300}\right)\right] \,,
\nonumber\\
&&J^C_{12}=\frac{1}{6\,\sqrt{v^2}\,((v \cdot u)^2 - v^2)}\left[(v \cdot u)\,\left(12\,C_{+1,1}^{210} + v^2\,\left(3\,v^2\,\left(C_{0,1}^{020} - C_{-1,1}^{110}\right)
\right.\right.\right.
\nonumber\\
&&\left.\left.\left.\quad + 6\,\left(C_{-1,0}^{110} + C_{0,1}^{120}\right) + 4\,C_{+1,2}^{210}\right)\right) \right.
+ (v \cdot u)^3\,\left(3\,v^2\,C_{-1,1}^{110} - 4\,C_{0,2}^{300}\right)
\nonumber\\
&&\left.\quad+6\,(v \cdot u)^2\,\left(2\,C_{0,1}^{210} - C_{+1,2}^{300} \right) \right.
\nonumber\\
&&\quad \left.
+ (v \cdot u)^2\, v^2\,\left(3\,C_{0,1}^{110} - 3\,C_{+1,2}^{200} + 12\,C_{0,2}^{210} -
      8\,C_{+1,3}^{300}\right)
\right.\nonumber\\
&&\left.\quad + (v ^2)^2\,(-3\,C_{0,1}^{110} + 6\,C_{+1,2}^{200} - 8\,C_{0,2}^{210} + 4\,C_{+1,3}^{300})\right] \,.
\end{eqnarray}

\newpage

\section*{Appendix G}

In this appendix we detail the form of the free-space loop matrices,
$J^V_{[ij]}(v,u)$ and $ \Delta J^V_{[ij]}(v,u)$ of (\ref{final-loop-renormalized}), in terms
of the six non-vanishing master functions, $J_{i}^V(w)$ introduced in (\ref{impose-generalization}). The results are
presented in two steps. First we introduce 13 intermediate loop functions $J_i^H(v,u)$ with:
\begin{eqnarray}
&&J_0^H(v,u)= J_0^V \,, \qquad \qquad \quad
 J_1^H(v,u) =
-\Sigma_V\,\frac{(v \cdot u)}{\sqrt{v^2}}\,J_0^V +\frac{(v \cdot w)}{\sqrt{v^2}\,\sqrt{w^2}}\,J_1^V\,,
\nonumber\\
&& J_2^H(v,u) = \Sigma_V\,\sqrt{\frac{(v\cdot u)^2-v^2}{v^2}}\,J_0^V
- \frac{(X \cdot w)}{\sqrt{w^2}}\,J_1^V\,, \qquad \;\;J_3^H(v,u) = J_3^V\,,
\nonumber\\
&& J_4^H(v,u) = \Sigma_V^2\,\frac{(v\cdot u)^2}{v^2}\,J_0^V
-2\,\Sigma_V\,\frac{(v\cdot u)\,(v \cdot w)}{v^2\,\sqrt{w^2}}\,J_1^V
\nonumber\\
&& \qquad \qquad +\, \frac{v^2\,w^2-(v\cdot w)^2}{v^2\,w^2}\,J_3^V+ \frac{(v \cdot w)^2}{v^2\,w^2}\,J_4^V\,,
\nonumber\\
&& J_5^H(v,u) = \Sigma_V^2\,\frac{(v \cdot u)^2-v^2}{v^2}\,J_0^V -2\,\Sigma_V\,
\sqrt{\frac{(v \cdot u)^2-v^2}{v^2}}\,\frac{(X \cdot w)}{\sqrt{w^2}}\,J_1^V
\nonumber\\
&& \qquad \qquad -\,\left(1+ \frac{(X\cdot w)^2}{w^2} \right)J_3^V
+\frac{(X\cdot w)^2}{w^2}\, J_4^V\, ,
\nonumber\\
&& J_6^H(v,u) = -\Sigma_V^2\,\frac{(v \cdot u )}{\sqrt{v^2}}\,
\sqrt{\frac{(v \cdot u)^2-v^2}{v^2}}\,J_0^V+ \frac{(X \cdot w)\,(v \cdot w)}{\sqrt{v^2}\,w^2}\,(J_3^V-J_4^V)
\nonumber\\
&& \qquad \qquad -\,\Sigma_V\,\frac{v^2\,((v \cdot w)+(v \cdot u)\,(w\cdot u))-2\,(v \cdot u)^2(v \cdot w)}{\sqrt{(v\cdot u)^2-v^2}\,\sqrt{w^2}\,v^2}\,J_1^V
\,,
\nonumber\\
&& J_7^H(v,u) = -\Sigma_V\,\frac{(v \cdot u)}{\sqrt{v^2}}\,J_3^V + \frac{(v \cdot w)}{\sqrt{v^2}\,\sqrt{w^2}}\,J_7^V \,,
\nonumber\\
&& J_8^H(v,u) =\Sigma_V\,\sqrt{\frac{(v \cdot u)^2-v^2}{v^2}}\,J_3^V  -\frac{(X \cdot w)}{\sqrt{w^2}}\,J^V_7\,,
\nonumber\\
&& J_9^H(v,u) = -\Sigma_V^3\,\frac{(v \cdot u)^3}{v^2\,\sqrt{v^2}}\,J_0^V
-3\,\Sigma_V^2\,\frac{(v\cdot u)^2}{v^2}\,J_1^H  -3\,\Sigma_V\,\frac{(v \cdot u)}{\sqrt{v^2}}\,J_4^H
\nonumber\\
&& \qquad \qquad - 3\,\frac{(v \cdot w)}{\sqrt{w^2}\,\sqrt{v^2}}\,\frac{(v \cdot w)^2-v^2\,w^2}{v^2\,w^2} J_7^V
+\frac{(v \cdot w)^3}{\sqrt{w^2}^3\,\sqrt{v^2}^3}\,J_9^V\,,
\nonumber\\
&& J_{10}^H(v,u) = -\Sigma_V^2\,\frac{(v \cdot u)^2}{v^2}\,J_2^H
- 2\,\Sigma_V\,\frac{(v \cdot u)}{\sqrt{v^2}}\,J_6^H
\nonumber\\
&& \qquad \qquad+\Sigma_V\,\sqrt{\frac{(v \cdot u)^2-v^2}{v^2}}\,\Big(
\frac{v^2\,w^2-(v \cdot w)^2}{v^2\,w^2}\,J_3^V +
\frac{(v \cdot w)^2}{v^2\,w^2}\,J_4^V \Big)
\nonumber\\
&& \qquad \qquad + \frac{3\,(v \cdot w)^2-v^2\,w^2}{v^2\,w^2}\,\frac{(X \cdot w)}{\sqrt{w^2}}\,J_7^V
- \frac{(v \cdot w)^2}{v^2\,w^2}\,\frac{(X \cdot w)}{\sqrt{w^2}}\,J_9^V\,,
\nonumber\\
&& J_{11}^H(v,u) =\Sigma_V^3\,\left(\frac{(v \cdot u)^2-v^2}{v^2} \right)^{3/2}J_0^V
 -3\,\Sigma_V^2\,\frac{(v \cdot u)^2-v^2}{v^2}\,J_2^H
\nonumber\\
&& \qquad \qquad+3\,\Sigma_V\,\sqrt{\frac{(v \cdot u)^2-v^2}{v^2}}\,J_5^H
+ 3\,\left(1+ \frac{(X \cdot w)^2}{w^2} \right)\,\frac{(X \cdot w)}{\sqrt{w^2}}\,J_7^V
\nonumber\\
&& \qquad \qquad -\frac{(X \cdot w)^3}{w^2\,\sqrt{w^2}}\,J_9^V \,,
\nonumber\\
&& J_{12}^H(v,u) = 2\,\Sigma_V\,
\sqrt{\frac{(v \cdot u)^2-v^2}{v^2}}\,J_6^H - \Sigma_V^2\,\frac{(v \cdot u)^2-v^2}{v^2}\,J_1^H
\nonumber\\
&& \qquad \qquad +\Sigma_V\,\frac{(v \cdot u)}{\sqrt{v^2}}\,\left(
\frac{w^2+(X \cdot w)^2}{w^2}\,J_3^V-\frac{ (X \cdot w)^2}{w^2}\,J_4^V \right)
\nonumber\\
&& \qquad \qquad + \frac{(v \cdot w)\,(X\cdot w)^2}{\sqrt{v^2}\,w^2\,\sqrt{w^2}}\,J_9^V
- \left(1+ 3\,\frac{(X \cdot w)^2}{w^2} \right)\,\frac{(v\cdot w)}{\sqrt{v^2}\,\sqrt{w^2}}\,J_7^V\,,
\end{eqnarray}
where
\begin{eqnarray}
X_\mu =
\frac{(v\cdot u)\,v_\mu-v^2\,u_\mu}{v^2\,\sqrt{(v \cdot
u)^2/v^2-1}}\,.
\end{eqnarray}
For $J^V_{[ij]}(v,u)$ the algebra of Appendix D applies
with the substitution $J_i\to J_i^H$ and $\bar m_N \to m_N$. The objects $ \Delta J^V_{[ij]}(v,u)$
are given by
\begin{eqnarray}
&&\Delta J^{(q)}_{[11]}=-\Delta J^{(q)}_{[22]}=\frac{(v \cdot u)}{\sqrt{v^2}}\,J^H_3\;, \qquad
 \Delta J^{(q)}_{[12]}= i\,\sqrt{\frac{(v \cdot u)^2}{v^2}-1}\,J^H_3\,,
\nonumber\\
&&\Delta J^{(p)}_{[11]} = -\Delta J^{(p)}_{[22]}= \frac{(v \cdot u)}{\sqrt{v^2}}\,J^H_{0} \;, \qquad
 \Delta J^{(p)}_{[12]} = i\,\sqrt{\frac{(v \cdot u)^2}{v^2}-1}\,J^H_0\,,
\nonumber\\
&&\Delta J^{(p)}_{[13]}=\Delta J^{(p)}_{[24]}= -\frac{1}{\sqrt{2}}\,\Delta J^{(p)}_{[18]}=-\frac{1}{\sqrt{2}}\,\Delta J^{(p)}_{[27]}=\frac{-1}{\sqrt{3}}\,
\sqrt{\frac{(v \cdot u)^2}{v^2}-1}\,J^H_2\,,\qquad
\nonumber\\
&&\Delta J^{(p)}_{[16]}=\Delta J^{(p)}_{[25]}=i\,\sqrt{\frac{(v \cdot u)^2}{v^2}-1}\,\Big(\sqrt{v^2}\,J^H_0-J^H_1\Big) ,
\nonumber\\
&&\Delta J^{(p)}_{[15]}=-\Delta J^{(p)}_{[26]}=\frac{(v \cdot u)}{\sqrt{v^2}}\,\Big(\sqrt{v^2}\,J^H_0-J^H_1 \Big)\;, \qquad
\nonumber\\
&&\Delta J^{(p)}_{[14]}=-\Delta J^{(p)}_{[23]}= \frac{1}{\sqrt{2}}\,\Delta J^{(p)}_{[28]}=-\frac{1}{\sqrt{2}}\,\Delta J^{(p)}_{[17]}
=\frac{i}{\sqrt{3}}\,\frac{(v \cdot u)}{\sqrt{v^2}}\,J^H_2\,,
\label{delta-k-a}
\end{eqnarray}
and
\begin{eqnarray}
&& \Delta J^{(p)}_{[33]}=-\Delta J^{(p)}_{[44]}= -\frac{1}{3}\,\frac{(v \cdot u)}{\sqrt{v^2}}\,\Big(2\,J^H_3-J^H_5 \Big)
\,,
\nonumber\\
&&
\Delta J^{(p)}_{[55]}=-\Delta J^{(p)}_{[66]}=\frac{(v \cdot u)}{\sqrt{v^2}}\,\Big(v^2\,J^H_0-2\,\sqrt{v^2}\,J^H_1+J^H_4 \Big)\,,
\nonumber\\
&& \Delta J^{(p)}_{[77]}=-\Delta J^{(p)}_{[88]}=\frac{1}{3}\,\frac{(v \cdot u)}{\sqrt{v^2}}\,\Big(J^H_3-2\,J^H_5 \Big)\,,
\nonumber\\
&& \Delta J^{(p)}_{[35]}= \Delta J^{(p)}_{[46]}=\frac{-1}{\sqrt{2}}\,\Delta J^{(p)}_{[58]}=
\frac{-1}{\sqrt{2}}\,\Delta J^{(p)}_{[67]}
\nonumber\\
&& \qquad \quad\,=-\frac{1}{\sqrt{3}}\,\sqrt{\frac{(v \cdot u)^2}{v^2}-1}\,
\Big(\sqrt{v^2}\,J^H_2-J^H_6 \Big) \,,
\nonumber\\
&&\Delta J^{(p)}_{[37]}=
\Delta J^{(p)}_{[48]}=-i\,\frac{\sqrt{2}}{3}\,\sqrt{\frac{(v \cdot u)^2}{v^2}-1}\Big(J^H_3-J^H_5\Big)\,,
\nonumber\\
&&\Delta J^{(p)}_{[57]}=-\Delta J^{(p)}_{[68]}=-i\,\sqrt{\frac{2}{3}}\,\frac{(v \cdot u)}{\sqrt{v^2}}\,\Big(\sqrt{v^2}\,J^H_2-J^H_{6}\Big)\,
\,,\nonumber\\
&&\Delta J^{(p)}_{[34]}=-\frac{i}{3}\,\sqrt{\frac{(v \cdot u)^2}{v^2}-1}\,\Big(2\,J^H_3+J^H_{5}\Big)\,,\quad
\nonumber\\
&&
\Delta J^{(p)}_{[78]}=-\frac{i}{3}\,\sqrt{\frac{(v \cdot u)^2}{v^2}-1}\,\Big(J^H_3+2\,J^H_{5}\Big)\,,
\nonumber\\
&&\Delta J^{(p)}_{[36]}=-
\Delta J^{(p)}_{[45]}=-\frac{i}{\sqrt{3}}\,\frac{(v \cdot u)}{\sqrt{v^2}}\,\Big(\sqrt{v^2}\,J^H_2-J^H_{6}\Big)\,,
\nonumber\\
&&\Delta J^{(p)}_{[38]}=-\Delta J^{(p)}_{[47]}=-\frac{\sqrt{2}}{3}\,\frac{(v \cdot u)}{\sqrt{v^2}}\,\Big(J^H_3+J^H_{5}\Big)\,,
\nonumber\\
&&\Delta J^{(p)}_{[56]}= i\,\sqrt{\frac{(v \cdot u)^2}{v^2}-1}\,
\Big(v^2\,J^H_0-2\,\sqrt{v^2}\,J^H_1+J^H_4 \Big)
\,. \label{delta-k-b}
\end{eqnarray}

\end{document}